\documentclass[12pt,a4paper]{article}

\usepackage{authblk}

\usepackage{a4wide}

\usepackage{oxford2}

\usepackage{color}
\usepackage{tikz}

\usepackage{a4wide}
\usepackage{textcomp}

\usepackage[english]{babel}

\usepackage{amsmath, amssymb, amsfonts}
\usepackage{fancyvrb}
\usepackage{graphicx}

\usepackage[pagestyles]{titlesec}
\newpagestyle{copyright}{
\setfoot[][][\footnotesize\copyright~B.~Wouters (UvA), 2019 \hfill \mbox{}] 
{}{}{\hfill \mbox{} \footnotesize\copyright~B.~Wouters (UvA), 2019}
}

\numberwithin{equation}{section}

\usepackage{transparent}

\usepackage{centernot}

\usepackage{diagbox}

\usepackage{bm}

\newcommand{\dd}[1]{\,\text{d}{#1}}

\newcommand{\Ex}[1]{\text{E}\!\left[{#1}\right]}

\newcommand{\Var}[1]{\text{Var}\!\left[{#1}\right]}

\usepackage{color}

\VerbatimFootnotes

\usepackage[hyperfootnotes=false, hidelinks]{hyperref}

\newcommand{\be}{{\bf e}}

\newcommand{\bvarepsilon}{{\bm{\varepsilon}}}

\newcommand{\bphi}{{\bm{\phi}}}
\newcommand{\bpsi}{{\bm{\psi}}}

\newcommand{\bSigma}{{\bm{\Sigma}}}
\newcommand{\bxi}{{\bm{\xi}}}
\newcommand{\boldeta}{{\bm{\eta}}}
\newcommand{\bchi}{{\bm{\chi}}}

\usepackage{mathtools}

\newcommand{\mise}[1]{\textbf{MISE}_\text{{#1}}}

\usepackage{caption}
\title{Noise reduction for functional time series}

\author{Cees Diks}
\author{Bram Wouters}
\affil{University of Amsterdam}

\begin{document}

\maketitle

\begin{abstract}
\noindent A novel method for noise reduction in the setting of curve time series with error contamination is proposed, based on extending the framework of functional principal component analysis (FPCA). We employ the underlying, finite-dimensional dynamics of the functional time series to separate the serially dependent dynamical part of the observed curves from the noise. Upon identifying the subspaces of the signal and idiosyncratic components, we construct a projection of the observed curve time series along the noise subspace, resulting in an estimate of the underlying denoised curves. This projection is optimal in the sense that it minimizes the mean integrated squared error. By applying our method to similated and real data, we show the denoising estimator is consistent and outperforms existing denoising techniques. Furthermore, we show it can be used as a pre-processing step to improve forecasting.
\end{abstract}

\newpage
\tableofcontents

\newpage
\section{Introduction} \label{sec:introduction}
Due to an abundance of data in our modern day and age, curve time series, also known as functional time series, are increasingly encountered across various disciplines of society~\cite{Bosq.2000,Hörmann.2010,Hörmann.2012,Panateros.2013,Aue.2015}.
Curve time series can either arise as functions observed at consecutive discrete moments in time, such as curves describing the term structure of interest rates~\cite{Hays.2012,Caldeira.2017,Andreasen.2019,Sen.2019}, return density curves~\cite{Bathia.2010} and near-infrared spectroscopy data~\cite{Yang.2022}, or by splitting an underlying continuous-time process into consecutive equal-length time segments, such as periodic weather record charts~\cite{Shang.2011}, intraday energy consumption curves~\cite{Cho.2013} and hourly concentration patterns of pollutants~\cite{Hörmann.2015}. 
The observed curve time series generally consist of a dynamical part, which is characterised by serial dependence across different curves in the time series, and a white noise part. These parts are latent, i.e.~not separately observable, challenging the identification and modelling of the dynamical part of curve (or more generally, high-dimensional) time series, which currently is an active area of research \cite{Qin.2020,Gao.2021,Chen.2022,Cubadda.2022,Dong.2022,Yang.2022,Qin.2022,Chang.2023}.

In this paper we propose a novel pre-processing methodology that filters out the noise from a functional times series, thereby giving access to the dynamics. Following \citetext{Bathia.2010} we consider a univariate curve time series
\begin{equation}
Y_t(u) = X_t(u) + \varepsilon_t(u), \label{eq:curve_time_series}
\end{equation}
where $t=1,\ldots,n$ labels the time steps and $u\in \mathcal{I},$ with $\mathcal{I}$ being a bounded interval on which the curves are defined. The nature of this interval (e.g., temporal, spatial) depends on the context in which the time series arises. Only the curves $Y_t(\cdot) \in L^2(\mathcal{I})$ can be observed and 
are thought of as consisting of a sum of an unobservable signal curve $X_t(\cdot)$ and an unobservable noise curve $\varepsilon_t(\cdot)$ according to \eqref{eq:curve_time_series}. By definition, the signal curve consists of the part of $Y_t(\cdot)$ that exhibits serial correlation and is in that sense dynamical. The remaining part of $Y_t(\cdot)$, which by assumption has zero autocorrelation, is associated with the noise curve. This so-called error contamination accounts for several potential sources of noise. For processes containing a component without serial dependence, this non-dynamical part will be attributed to the noise curves. If the signal is observed imperfectly, idiosyncratic measurement errors are also part of the noise curves. One can think of, for instance, experimental errors or the use of discrete grids giving rise to numerical round-off errors~\cite{Bathia.2010}. Another example is when an observed signal requires estimation, thereby introducing estimation error.

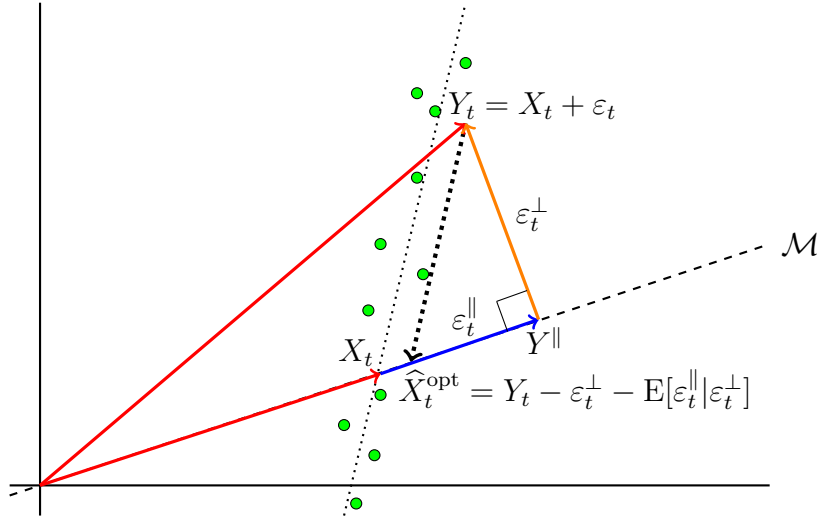
\begin{figure}[h]
\begin{center}
 \begin{tikzpicture}[scale=0.8]
\draw[-,thick] (-0.5,0)--(12,0); 
\draw[-,thick] (0,-0.5)--(0,8); 
\draw[black,fill=green] (7,7) circle (.5ex);
\draw[black,fill=green] (6.2,6.5) circle (.5ex);
\draw[black,fill=green] (6.5,6.2) circle (.5ex);
\draw[black,fill=green] (6.2,5.1) circle (.5ex);
\draw[black,fill=green] (5.6,4) circle (.5ex);
\draw[black,fill=green] (6.3,3.5) circle (.5ex);
\draw[black,fill=green] (5.4,2.9) circle (.5ex);
\draw[black,fill=green] (5.6,1.5) circle (.5ex);
\draw[black,fill=green] (5,1) circle (.5ex);
\draw[black,fill=green] (5.5,0.5) circle (.5ex);
\draw[black,fill=green] (5.2,-0.3) circle (.5ex);
\draw[dashed, thick] (-0.5,-0.167)--(12,4) node[right]{$\mathcal{M}$};
\draw[dotted, thick] (5.,-0.5)--(7,8); 
\draw[->, dotted, ultra thick] (7,6)--(6.1,2.05);
\node at (5.2,2.2) {$X_t$};
\node at (8.3,2.4) {$Y^{\|}$};
\node at (8.1,6.3) {$Y_t = X_t + \varepsilon_t$};
\node at (7,2.8) {$\varepsilon^{\|}_t$};
\node at (8.1,4.5) {$\varepsilon^{\perp}_t$};
\node at (8.8,1.6) {$\widehat{X}_t^\text{opt} = Y_t - \varepsilon^{\perp}_t - \text{E}[\varepsilon^{\|}_t|\varepsilon^{\perp}_t]$};
\draw[->,red, very thick] (0,0)--(5.6,1.85);
\draw[->, red, very thick] (0,0)--(7,6);
\draw[->,blue, very thick] (5.6,1.85)--(8.20,2.75);
\draw[->,orange, very thick] (8.20,2.75)--(7,6);
\draw[-, thin] (7.51,3.05)--(8.01,3.23);
\draw[-, thin] (7.51,3.05)--(7.69,2.55);
\end{tikzpicture}
\end{center}
 \caption{A sketch of the main idea of MISE-optimal denoising, with a 1-dimensional dynamical space $\mathcal{M}$ and a 2-dimensional noise space $\mathcal{M}_\varepsilon.$ In the expression for the reconstructed noise $\widehat{X}_t^\text{opt},$ not only the perpendicular noise is subtracted from the observed time series, but also the conditional expectation of the parallel noise component.}
 \label{fig:sketch_method}
\end{figure}

A common approach to the analysis of curve time series is by functional principal component analysis (FPCA) \cite{Dauxois.1982,Jones.1992}, where the signal curves are decomposed in terms of $\mathcal{M} = \text{span}\{ \phi_1(u), \ldots ,\phi_d(u) \},$ a low-dimensional space of curves which we will call the dynamical space or signal space. This decomposition is given by
\begin{equation}
X_t(u) - \mu(u) = \sum_{i=1}^d \xi_{ti} \phi_i(u) , \label{eq:finite_d_assumption}
\end{equation}
where $\mu(u):=\Ex{X_t(u)}$ and the random variables $\xi_{ti}$ constitute a (by assumption) stationary vector-valued time series. From the perspective of FPCA they can be called ``principal components''. An alternative perspective is provided by high-dimensional factor models (see e.g.~\citeauthor{Hallin.2023}, \citeyear{Hallin.2023} and \citeauthor{Tavakoli.2023}, \citeyear{Tavakoli.2023}), in the context of which $\xi_{ti}$ would be called ``(factor) loadings''.

\citetext{Bathia.2010} used the dynamical properties (i.e.\ serial dependence) of the signal curves to estimate $d$ as well as to find an orthonormal basis of $\mathcal{M}.$ This is a non-trivial task, as a naive implementation of FPCA would lead to an asymptotically biased estimate due to a non-neglible contribution from the noise covariance $\Sigma_\varepsilon(u,v) := \text{Cov} \left[ \varepsilon_t(u), \varepsilon_t(v) \right].$ We will refer to this method as ``dynamical functional principal component analysis'' (DFPCA). Building on \citetext{Bathia.2010}, our two main contributions are:
\begin{itemize}
\item  under mild assumptions about the time series of the principal components $\xi_{ti}$, we are able to estimate the noise covariance $\Sigma_\varepsilon(u,v).$ We use this to identify a finite-dimensional structure of the space of noise curve $\mathcal{M}_\varepsilon,$ analogously to FPCA-based estimation of the dynamical space $\mathcal{M}$;
\item using this structure, we decompose the noise curves $\varepsilon_t(u) = \varepsilon_t^\parallel(u) + \varepsilon_t^\perp(u)$ in a component parallel to the dynamical space $\mathcal{M}$ and a perpendicular component. Given an estimate of $\mathcal{M},$ we can observe the perpendicular noise $\varepsilon_t^\perp(\cdot)$ component via orthogonal projection of the observed curves onto the dynamical space, and use the knowledge of the noise covariance $\Sigma_{\varepsilon}(\cdot,\cdot)$ to fit a linear regression model for the conditional expectation $\text{E}\big[\varepsilon_t^\parallel(\cdot) \big| \varepsilon_t^\perp(\cdot)\big].$ This in turn is used to reconstruct the unobservable signal curves $X_t(\cdot)$ in a way that minimizes the mean integrated squared error (MISE) between the original signal curve and its reconstruction. We call this method, which is graphically illustrated in Figure~\ref{fig:sketch_method}, MISE-optimal denoising. 
\end{itemize}
In a simulation study we show that MISE-optimal denoising is consistent, in the sense that the obtained MISE converges to its theoretical minimum as the time series length increases. Furthermore, we show that using a MISE-optimal denoised signal for forecasting outperforms other forecasting approaches. We also succesfully apply MISE-optimal denoising to an empirical dataset of intraday temperature curves.

We emphasize that the proposed noise reduction method should be viewed as a pre-processing step, using only mild assumptions about the underlying dynamical model. After removing (part of) the noise from the observed curve time series, the remaining denoised signal curves can be used in, e.g., existing modelling and forecasting techniques. The idea is that the noise reduction step helps improve the estimation of the dynamics and as a result also improves forecasting performance.

This paper is organized as follows. After a discussion of the related literature, Section~\ref{sec:methodology} builds on DFPCA~\citetext{Bathia.2010} to identify the structure of the noise space $\mathcal{M}_\varepsilon$ and to derive an estimator for MISE-optimal denoising. Section~\ref{sec:estimation} discusses a number of practical issues related to estimation in the context of MISE-optimal denoising. In Section~\ref{sec:simulation} MISE-optimal denoising is applied to simulated data and its performance is scrutinized in a variety of ways. An application to empirical data is given in Section~\ref{sec:weather_data}.

\subsection{Related literature}
Dimension reduction in functional data analysis has a long history, documented in a vast body of literature; see, e.g., \citetext{Ramsay.2005} and references therein. In the presence of serial dependence we talk about curve or functional time series \cite{Bosq.2000,Hörmann.2010,Hörmann.2012,Panateros.2013,Li.2020}. For early literature about dimension reduction in this context, see \cite{Bosq.2000,Ferraty.2006}. A major advancement in this field came with \citetext{Bathia.2010}, which identifies the ``dimensionality of a curve time series'', meaning the dimension and basis vectors of (what we call) the signal space. This is also the main inspiration of this paper. Ideas similar to \citetext{Bathia.2010} have also been applied to (latent) factor models for high-dimensional time series~\cite{Lam.2011,Lam.2012}, which are closely related to the curve time series setting we study. \citetext{Cubadda.2022} studied conditions for the existence of such a signal-noise decomposition in the context of high-dimensional time series. Other attempts at separating signal from noise have as disadvantage that they make strong additional assumptions about the structure of the noise \cite{Yao.2005,Hall.2006,Descary.2019}.

A related form of denoising was considered by \citetext{Dong.2022}, who project a finite-dimensional time series onto a lower-dimensional time series, where the projection is determined by minimizing a forecast error.
\citetext{Chen.2022} developed a method for functional linear regression, rather than noise reduc in a time series setting, using similar elements from \citetext{Bathia.2010} as we do. In order fit a (scalar) response $R_t$ with respect to a functional predictor $X_t(\cdot),$ corresponding to the signal curves in our setup, a slope function $\beta(\cdot)$ must be estimated such that
\begin{equation}
R_t = \int_\mathcal{I} X_t(u) \beta(u) \dd u + \varepsilon_t',
\end{equation}
where $\varepsilon_t'$ is some idiosyncrasy term. The main similarity lies in their usage of the fact that the autocovariance with nonzero lag of the signal curves $X_t(\cdot)$ is equal to the autocovariance of the observed curves $Y_t(\cdot).$ They use this to define a generalized method-of-moments estimator for the slope function of functional linear regression. \citetext{Chang.2023} put this approach in a broader framework and generalised it to multivariate functional time series.

\section{Methodology} \label{sec:methodology}
In this section we present our main theoretical results: a method to identify the structure of the noise space and a MISE-optimal denoising algorithm. We start with a brief review of DFPCA, as this is the starting point of our main contributions. To improve readability, technical details are often deferred to appendices.

\subsection{A review of DFPCA} \label{sec:DFPCA}
A naive approach to finding estimates of the dimension $d$ and a basis of the dynamical space $\mathcal{M}$ would consist of performing an eigensystem analysis of
\begin{equation}
\Sigma_{X}(u,v) := \text{Cov} \left[ X_t(u), X_t(v) \right] = \sum_{i=1}^d \lambda_i \phi_i(u) \phi_i(v), \label{eq:spectral_decomp_SigmaX}
\end{equation}
where \eqref{eq:finite_d_assumption} was used, as well as the following properties of the principal components $\xi_{ti}$ as proven by the Karhunen-Lo\`eve theorem:
\begin{equation}
\Ex{\xi_{ti}} = 0, \qquad \Var{\xi_{ti}} = \lambda_i, \quad \text{and} \quad \Ex{\xi_{ti}\xi_{tj}} = 0 \quad  \text{if } i \neq j, \label{eq:KarhunenLoeve_properties}
\end{equation}
which hold for all $t=1,2\ldots,n.$ Note that by Mercer's theorem all eigenvalues in \eqref{eq:spectral_decomp_SigmaX} are non-negative and for convenience we assume $\lambda_1 \geq \lambda_2 \geq \ldots \geq \lambda_d > 0.$ The problem with this approach is that $\Sigma_{X}(\cdot,\cdot)$ cannot be estimated without bias, since the curves $X_t(\cdot)$ cannot be observed directly and $\Sigma_{Y}(u,v) := \text{Cov} \left[ Y_t(u), Y_{t}(v) \right] = \Sigma_{X}(u,v) + \Sigma_{\varepsilon}(u,v).$ Using the observable curves $Y_t(\cdot)$ to estimate $\Sigma_{X}(\cdot,\cdot)$ leads to a bias, in particular for relatively large noise.  \citetext{Hall.2006} circumvented this problem by assuming independence of the curves $Y_1(\cdot), \ldots, Y_n(\cdot),$ as well as vanishing noise curves $\varepsilon_t(\cdot)$ in the limit of infinite sample size $n.$ It should be noted that this method does not use the serial dependence of the curve time series.

\citetext{Bathia.2010} proposed an innovative method for estimating $d$ and a basis for $\mathcal{M}$, based on the fact that the noise curves (by definition) do not exhibit any serial dependence. In other words, the autocovariance of the noise for nonzero lag is zero, $\text{Cov} \left[ \varepsilon_t(u), \varepsilon_{t+k}(v) \right] =0$ for any $k \neq 0.$ This implies that the lag-$k$ autocovariance of the signal curves, $M_k(u,v) := \text{Cov} \left[ X_t(u), X_{t+k}(v) \right],$ equals the lag-$k$ autocovariance of the observed curves:
\begin{equation}
M_k(u,v) = \text{Cov} \left[ Y_t(u), Y_{t+k}(v) \right], \label{eq:Bathia_starting_point}
\end{equation}
for any $k \neq 0.$ This is crucial, as now the observed curves $Y_t(\cdot)$ can be used to estimate the autocovariance of the signal curves $X_t(\cdot)$ for nonzero lag. \citetext{Bathia.2010} then exploit the fact that, under certain assumptions (see Appendix
\ref{app:background}
for details), the $d$-dimensional eigenspace of any operator
\begin{equation}
K(u,v) := \sum_{\ell = 1}^q c_\ell N_\ell (u,v) , \qquad \text{with}\quad N_k (u,v) : = \int_\mathcal{I} M_k(u,z) M_k(v,z) \dd z, \label{eq:K_definition}
\end{equation}
coincides with the dynamical space $\mathcal{M}.$ The coefficients $c_\ell$ can be chosen (almost) arbitrarily, provided at least one of them is nonzero. The various operators $K(\cdot,\cdot)$ that are obtained by different choices of the coefficients $c_\ell$, give different sets of eigenfunctions, but each set of eigenfunctions spans the same dynamical space $\mathcal{M}.$ \citetext{Bathia.2010} choose $q=5$ and $c_{\ell} = 1$ for $\ell = 1, \ldots, 5$.
We refer to their approach to finding $d$ and a basis for $\mathcal{M}$ as DFPCA, as it is using the dynamical properties encoded in the nonzero-lag autocovariances $M_{k\neq 0}(\cdot,\cdot)$. For more details about DFPCA, see Appendix~\ref{app:background}

\subsection{The structure of the noise space} \label{sec:finding_noise_covariance}
This section focuses primarily on the noise covariance $\Sigma_{\varepsilon}(\cdot,\cdot).$ Knowledge of this can provide insights in the structure of the noise via the FPCA framework. To be more concrete, it might enable us to identify a relatively small number of functions that account for most of the noise present in the curve time series. Because the noise curves $\varepsilon_t(\cdot)$ are not observable, it is not possible to directly estimate $\Sigma_{\varepsilon}(\cdot,\cdot).$ Since $\Sigma_{\varepsilon}(u,v) = \Sigma_{Y}(u,v) - \Sigma_{X}(u,v)$ and $\Sigma_{Y}(\cdot,\cdot)$ can be estimated directly, the question is whether we can get access to the covariance of the signal curves $\Sigma_{X}(\cdot,\cdot).$

DFPCA provides an alternative basis $\psi_1(\cdot), \ldots ,\psi_d(\cdot)$ of the signal curve space $\mathcal{M}$ that, unlike the basis of \eqref{eq:finite_d_assumption}, can be estimated consistently. With respect to this alternative basis the demeaned signal curves can be expressed as,
\begin{equation} \label{eq:two_expansions}
X_t(u) - \mu(u)  =  \sum_{i=1}^d \xi_{ti} \phi_i(u) =  \sum_{i=1}^d \eta_{ti} \psi_i(u), \quad \text{where}\quad \eta_{ti} = \int_\mathcal{I} \left[ X_t(u) - \mu(u)  \right] \psi_i(u) \dd u,
\end{equation}
implying that
\begin{equation}
M_k(u,v) = \bphi(u)^T {\Sigma}_k \, \bphi(v) = \bpsi(u)^T {\Sigma}_k^{(\eta)} \, \bpsi(v), \label{eq:Mk_two_forms}
\end{equation}
for any $k \in \mathbb{Z},$ where ${\Sigma}_k := \Ex{\bxi_t \bxi_{t+k}^T}$ and ${\Sigma}_k^{(\eta)} :=\Ex{\boldeta_t \boldeta_{t+k}^T}$ are the finite-dimensional variance-covariance matrices of the principal components of the two respective bases of $\mathcal{M}.$ Note that boldface greek letters denote vectors, e.g.~$\bphi(u) := \big( \phi_1(u), \ldots ,\phi_d(u) \big)$ and $\bxi_t := \big( \xi_{t1}, \ldots, \xi_{td}\big).$ This in particular means that $\Sigma_{X}(u,v) = M_0(u,v) = \bpsi(u)^T {\Sigma}_0^{(\eta)} \, \bpsi(v).$ Access to the matrix ${\Sigma}_0^{(\eta)}$ is what separates us from being able to estimate the covariance of the signal curves $\Sigma_{X}(\cdot,\cdot).$

The sample versions of the principal components $\eta_{ti}$ defined in \eqref{eq:two_expansions} cannot be computed directly from the data, as the signal curves are unobservable. However, they have proxies
\begin{equation}
\chi_{ti} := \int_\mathcal{I} \left[ Y_t(u) - \mu(u)  \right] \psi_i(u) \dd u = \eta_{ti} + \varepsilon'_{ti} ,\qquad \text{where}\quad \varepsilon'_{ti} := \int_\mathcal{I} \varepsilon_t(u) \psi_i(u) \dd u,
 \label{eq:chi_definition}
\end{equation}
that can be computed given the observable curves $Y_t(\cdot).$ Conveniently, using the fact that the noise curves (by definition) do not have serial dependence, their covariance matrices are related via
\begin{equation}
\Sigma_k^{(\chi)} : = \text{Cov} \left[ \bchi_t, \bchi_{t+k} \right] = \left\{
\begin{matrix}
\Sigma_k^{(\eta)} + \text{Cov} \left[ \bvarepsilon'_t, \bvarepsilon'_t \right]  & \qquad \text{if }k=0, \\
\Sigma_k^{(\eta)} & \qquad \text{if }k\neq0.
\end{matrix}
\right.
\end{equation}
This means that the covariance matrices $\Sigma_k^{(\eta)}$ for any $k\neq0$ can be estimated through the proxy principal components $\chi_{ti}.$

To gain access to the lag-0 covariance matrix ${\Sigma}_0^{(\eta)},$ we wish to exploit the underlying dynamics. A simple assumption regarding the dynamics, is that the original principal components $\bxi_t$, defined in \eqref{eq:finite_d_assumption}, follow the dynamics of a lag-$p$ vector autoregressive process. We focus on the simplest case (lag-$1$) first, that is, the assumption that the time series of $\bxi_t$ is described by a VAR(1) model
\begin{equation}
\bxi_t = A \, \bxi_{t-1} + \be_t, \qquad \text{where} \quad \be_t \thicksim \text{IID} \left( \bm{0}, \Omega \right). \label{eq:VAR1_assumption}
\end{equation}
Note the absence of a constant term for the mean, since $\Ex{\xi_{ti}}=0.$ As a consequence of this assumption the random coefficients $\boldeta_t$ also follow a VAR(1) process. The lagged autocovariance matrices of a VAR($p$)-process time series are related through the so-called Yule-Walker equations (see Appendix~\ref{app:var} for details). For the VAR(1)-process of the principal components $\boldeta_t$ these Yule-Walker equations can be rearranged as
\begin{equation}
\Sigma_0^{(\eta)} = \Sigma_1^{(\eta)} \left(\Sigma_2^{(\eta)}\right)^{-1} \Sigma_1^{(\eta)} . \label{eq:YW_reconstruction}
\end{equation}
This is a crucial result, because this makes $\Sigma_0^{(\eta)}$ and thereby $\Sigma_{X}(\cdot,\cdot)$ accessible through the proxy principal components $\chi_{ti}$ defined in \eqref{eq:chi_definition}.

For VAR processes with lag order $p>1$ it is possible to generalize \eqref{eq:YW_reconstruction} and express $\Sigma_0^{(\eta)}$ in terms of $\Sigma_p^{(\eta)}, \Sigma_{p+1}^{(\eta)},\ldots, \Sigma_{p+p}^{(\eta)}.$ We have run the MISE-optimal denoising algorithm on the simulated data of section \ref{sec:simulation} for $p=2$ and $p=3.$ The results in terms of denoising performance and asymptotic behaviour were similar to the default case of $p=1$ and \eqref{eq:YW_reconstruction}. This indicates that our approach is insensitive to the assumption of an underlying VAR(1) process for the principal components $\bxi_t,$ even when this model is misspecified \cite{Dahlhaus.1996}. Throughout this paper we assume a VAR(1) process by default, as this choice is expected to suffer the least from small-sample estimation noise. For more details, see Appendix \ref{app:VAR_order_dependence}.

The above analysis is a key step for our approach, as this enables us to express the noise covariance operator $\Sigma_{\varepsilon}(u,v) = \Sigma_{Y}(u,v) - \Sigma_{X}(u,v),$ where $\Sigma_{X}(u,v) = \bpsi(u)^T \Sigma_0^{(\eta)} \, \bpsi(v)$, in terms of quantities that can be estimated. Since $\Sigma_{Y}(\cdot,\cdot),$ $\bpsi(\cdot)$ and $\Sigma_0^{(\eta)}$ can be estimated consistently, without bias from the noise curves $\varepsilon_t(\cdot)$, this provides a consistent estimator of the noise covariance operator $\Sigma_{\varepsilon}(\cdot,\cdot).$ 
This allows us to extend the FPCA framework to the noise space $\mathcal{M}_\varepsilon$. Assuming that the eigenspace of $\Sigma_{\varepsilon}(\cdot,\cdot)$ is of finite dimension, analoguous to $X_t(\cdot)$ also the noise curves $\varepsilon_t(\cdot)$ can (approximately) be expanded in terms of a finite number of eigenfunctions, in this case of the noise covariance $\Sigma_{\varepsilon}(\cdot,\cdot).$ The orthonormal eigenbasis $\phi_1^{(\varepsilon)}(\cdot), \ldots , \phi_{d_\varepsilon}^{(\varepsilon)}(\cdot)$ of $\Sigma_{\varepsilon}(\cdot,\cdot)$ forms a basis of the noise space $\mathcal{M}_\varepsilon$ such that
\begin{equation}
\varepsilon_t(u) = \sum_{i=1}^{d_\varepsilon} \xi_{ti}^{(\varepsilon)} \phi_i^{(\varepsilon)}(u), \label{eq:finite_d_eps_assumption}
\end{equation}
where the random variables $\xi_{ti}^{(\varepsilon)}$ have the same properties as in \eqref{eq:KarhunenLoeve_properties} due to the Karhunen-Lo\`eve theorem.

\subsection{MISE-optimal denoising} \label{sec:denoising}
Given an observed curve $Y_t(u) = X_t(u) + \varepsilon_t(u),$ the aim of denoising is to find a reconstruction $\widehat{X}_t^\text{den}(\cdot)$ of the unobservable signal curve $X_t(\cdot).$ An example of a denoising procedure is orthogonal denoising, where the observed curve $Y_t(\cdot)$ is projected orthogonally onto the dynamical space $\mathcal{M}.$ In other words, for orthogonal denoising the reconstructed signal curve is given by $\widehat{X}_t^\text{ortho}(u) = Y_t^\parallel(u) := (P_\mathcal{M}Y_t)(u),$ where $P_\mathcal{M}(u,v) = \bpsi(u) \cdot \bpsi(v)$ is the operator of the orthogonal projection onto $\mathcal{M}.$ Denoising performance can be measured in terms of the ``mean integrated square error'' (MISE)
\begin{equation} \label{eq:MSE_objective}
\mise{den} := \overline{\text{E}} \big[ ( X_t(\cdot) - \widehat{X}_t^\text{den}(\cdot) )^2 \big], \qquad\text{where}\quad
\overline{\text{E}} \left[ g(\cdot) \right] := \frac{1}{\Delta_\mathcal{I}} \int_\mathcal{I} \Ex{g(u)} \dd u ,
\end{equation}
in the case of a bounded interval of width $\Delta_\mathcal{I}.$ As we will see, $\mise{ortho}$ associated with $\widehat{X}_t^\text{ortho}(\cdot)$ is generally not optimal in the sense that it does not minimize the MISE.

Let us now ask the question of finding a MISE-optimal denoising procedure. Concretely, the goal is to find an operator $P: L^2(\mathcal{I}) \rightarrow L^2(\mathcal{I})$ such that the denoised curve ${X}_t^\text{opt}(u) := (PY_t)(u)$ minimizes the associated $\mise{opt}$ defined in \eqref{eq:MSE_objective}. Note the slight abuse of notation here, as \eqref{eq:MSE_objective} already contains an estimator and here we are searching for the operator that minimizes the MISE in \eqref{eq:MSE_objective}, which we will then call the MISE-optimal denoising estimator. In this paper the space of operators $P$ is restricted to (linear) projections onto the dynamical space, i.e.~$(Pf)(u) = f(u)$ for any function $f(\cdot) \in \mathcal{M}.$ Combined with the fact that a noise curve can be decomposed in a unique way in a part parallel to the dynamical space and a part orthogonal to the dynamical space, $\varepsilon_t(u) = \varepsilon_t^\parallel(u) + \varepsilon_t^\perp(u)$ where $\varepsilon_t^\parallel(\cdot) \in\mathcal{M}$ and $\varepsilon_t^\perp(\cdot)$ lies inside the orthogonal complement of $\mathcal{M},$ this implies that $X_t^\text{opt}(u) = Y_t^\parallel(u) + (P \varepsilon_t^\perp)(u),$ where $Y_t^\parallel(u) = X_t(u) + \varepsilon_t^\parallel(u).$

The challenge is now to find a projection such that the parallel part of the noise curve in $Y_t^\parallel(\cdot)$ is cancelled by $(P \varepsilon_t^\perp)(\cdot)$ in a MISE-optimal fashion. For this purpose, the latter is expanded as
\begin{equation}
(P \varepsilon_t^\perp)(u) = \sum_{i=1}^{d_\parallel} \sum_{j=1}^{d_\perp} \alpha_{ij} \, \varepsilon_{t,j}^\perp  \, \phi^\parallel_i (u), \label{eq:denoising_linear_model}
\end{equation}
in terms of an orthonormal eigenbasis $\phi^\parallel_1(\cdot), \phi^\parallel_2(\cdot), \ldots, \phi^\parallel_{d_\parallel}(\cdot)$ of $\mathcal{M}_\parallel,$ which is the space of noise curves $\varepsilon_t^\parallel(\cdot)$ parallel to $\mathcal{M}.$ Here, the $\varepsilon_{t,j}^\perp$ are the coordinates of the perpendicular part of the noise curves with respect to an orthonormal basis $\phi^\perp_1(\cdot), \phi^\perp_2(\cdot), \ldots, \phi^\perp_{d_\perp}(\cdot)$ of the space $\mathcal{M}_\perp$ of these perpendicular noise curves,
\begin{equation}
\varepsilon_t^\perp(u) = \sum_{j=1}^{d_\perp} \varepsilon_{t,j}^\perp \, \phi^\perp_j (u) .
\end{equation}
The finite dimensionality of the subspaces $\mathcal{M}_\|$ and $\mathcal{M}_\perp$ is ensured by applying the FPCA assumption to the noise space $\mathcal{M}_\varepsilon,$ as formulated in \eqref{eq:finite_d_eps_assumption}.

The idea behind \eqref{eq:denoising_linear_model} is that the perpendicular noise curves can be indirectly observed via $ \varepsilon_t^\perp(u) = Y_t(u) - Y_t^\parallel(u).$ They can therefore be used as a predictor, with the parallel noise curves $\varepsilon_t^\parallel(\cdot)$ as response variable. From this perspective, the parameters $\alpha_{ij}$ are the regression coefficients and they are determined by minimizing the MISE in \eqref{eq:MSE_objective}. We are essentially fitting a linear model for the conditional expectation $\text{E}\big[\varepsilon_t^\parallel(\cdot) \big| \varepsilon_t^\perp(\cdot)\big]$ that minimizes the MISE by exploiting the covariance between the perpendicular and parallel part of the noise curves. See Figure~\ref{fig:sketch_method} for a simplified illustration of this approach in the case of a bivariate time series. Note the absence of an intercept in \eqref{eq:denoising_linear_model}, because the noise curves have zero mean.

The result is a convex optimization problem with solution $\hat{\alpha} = - \Omega_{\parallel\perp} \left( \Omega_\perp \right)^{-1},$ where
\begin{subequations} \label{eq:Omega_perp_Omega_parallel_perp_def}
\begin{align}
\Omega_\perp &  := \int_\mathcal{I} \int_\mathcal{I}  \bphi^\perp(u)\left( \bphi^\perp(v)  \right)^T  \Sigma_{\varepsilon}(u,v) \dd u \dd v, \label{eq:Omega_perp}  \\
\Omega_{\parallel\perp} & := \int_\mathcal{I} \int_\mathcal{I}  \bphi^\parallel(u) \left(\bphi^\perp(v)  \right)^T \Sigma_\varepsilon(u,v) \dd u \dd v,
\end{align}
\end{subequations}
and where $\bphi^\parallel(u) := \big( \phi^\parallel_1(u), \phi^\parallel_2(u), \ldots, \phi^\parallel_{d_\parallel}(u) \big)^T$ and $\bphi^\perp(u) := \big( \phi^\perp_1(u), \phi^\perp_2(u), \ldots, \phi^\perp_{d_\perp}(u) \big)^T.$ The matrix $\Omega_\perp$ should be interpreted as the covariance of between the coordinates $\bvarepsilon_{t}^\perp:= \big(\varepsilon_{t,1}^\perp, \varepsilon_{t,2}^\perp, \ldots, \varepsilon_{t, d_\perp}^\perp \big)^T,$ whereas the matrix $\Omega_{\parallel\perp}$ is the covariance between the similarly defined coordinates $\bvarepsilon_{t}^\parallel := \big(\varepsilon_{t,1}^\parallel, \varepsilon_{t,2}^\parallel, \ldots, \varepsilon_{t, d_\parallel}^\parallel \big)^T$ and $\bvarepsilon_{t}^\perp.$ This leads to a MISE-optimal reconstruction of the signal curves $X_t(\cdot)$ given by
\begin{equation}
\widehat{X}_t^\text{opt}(u) = Y_t^\parallel(u) - \int_\mathcal{I} \left( \bphi^\parallel(u) \right)^T \Omega_{\parallel\perp} \left( \Omega_\perp \right)^{-1} \bphi^\perp(v) \left( Y_t(v) - Y_t^\parallel(v) \right)  \dd v. \label{eq:denoising_main_result}
\end{equation}
The above formula for the MISE-optimal denoising estimator of the signal curves is the main result of this paper. More details about the derivation of \eqref{eq:denoising_main_result} can be found in Appendix \ref{app:denoising}, as well as an insightful illustration of the denoising formula in the context of finite-dimensional vector spaces. 

Using this denoising approach, the minimum of the MISE is
\begin{equation} \label{eq:MISE_minimum}
\mise{opt}^\text{min} := 
\min_{\alpha \in \mathbb{R}^{(d_\| \times d_\perp)}} \mise{opt} = \text{Tr} \left[ \Omega_\parallel \right] - \text{Tr}\big[ \left(\Omega_\perp\right)^{-1} \left(\Omega_{\parallel\perp}\right)^T \Omega_{\parallel\perp} \big].
\end{equation}
It accounts for the irreducible components of the noise curves, which are present in the general case where the noise space and the dynamical space have overlap. In the special case of no overlap, i.e.~$\mathcal{M}_\varepsilon \cap \mathcal{M} = \emptyset,$ the minimum of the MISE is zero and perfect denoising is possible (at the population level). For more details, see Appendix~\ref{app:irreducible_noise}.

\subsection{The noise level} \label{sec:noise_level}
An important property of the curve time series~\eqref{eq:curve_time_series} is the noise level $\lambda \in [0,1]$, which we define as the relative size of the noise with respect to the observed curve variance,
\begin{equation}
\lambda : = \frac{\overline{\text{Var}}\left[{\varepsilon}_t(\cdot)\right]}{\overline{\text{Var}}\left[Y_t(\cdot)\right]}, \label{eq:relative_size_noise}
\end{equation}
where $\overline{\text{Var}}\left[g(\cdot)\right]$ is defined analoguous to $\overline{\text{E}} \left[ g(\cdot) \right]$ in \eqref{eq:MSE_objective}. Note that $\lambda$ is directly related to the (integrated) signal-to-noise ratio $\overline{\text{Var}}\left[{X}_t(\cdot)\right]/\overline{\text{Var}}\left[{\varepsilon}_t(\cdot)\right].$

A consequence of the proposed MISE-optimal denoising procedure is the ability to estimate the noise level. Since the noise cannot be observed directly, estimating its relative size is a nontrivial problem. Using the MISE-optimal denoised signal $\widehat{X}_t^\text{opt}(\cdot),$ one can show that
\begin{equation}
\lambda = \frac{\mise{opt}^\text{min} + \overline{\text{E}}\big[ ( Y_t(\cdot) - \widehat{X}_t^\text{opt}(\cdot) )^2 \big] }{\overline{\text{Var}}\left[{Y}_t(\cdot)\right]} , \label{eq:reconstruct_lambda}
\end{equation}
consisting of quantities that all can be estimated. Note that, since the analysis of section~\ref{sec:finding_noise_covariance} provides access to $\Sigma_\varepsilon(\cdot,\cdot),$ an alternative way of obtaining the noise level is through $\lambda = \text{Tr} \left[\Sigma_\varepsilon \right]/ \overline{\text{Var}}\left[{Y}_t(\cdot)\right].$
\newpage
\section{Estimation and consistency \label{sec:estimation}}
In this section we describe how MISE-optimal denoising can be applied in the default realistic scenario, in which an observed curve time series $\left\{ Y_t(u) \right\}_{t=1}^n$ of length $n$ is available. In particular, we describe how orthonormal bases of the subspaces $\mathcal{M},\, \mathcal{M}_\varepsilon,\, \mathcal{M}_\parallel$ and $\mathcal{M}_\perp$ can be estimated and how they can be used to define an estimated operator corresponding to MISE-optimal denoising. Furthermore, we define what we mean by consistency of a denoising method. This section focuses on the aspects of estimation that are specific for our denoising method, whereas more standard formulas for estimators can be found in Appendix~\ref{app:estimation}.

\subsection{Estimation of $\mathcal{M}$}
For the estimation of the dimension $d$ of the dynamical space and the basis functions $\psi_1(\cdot), \ldots, \psi_d(\cdot)$ we ollow the approach of \citetext{Bathia.2010}. The basis functions are the eigenfunctions of the estimate of the operator $K(\cdot,\cdot)$ defined in \eqref{eq:K_definition}. No smoothing methods are used to improve estimation precision. This choice has the advantage that MISE-optimal denoising as presented here does not make any assumptions about the smoothness of the curves of the functional time series and can therefore also be applied to vector-valued high-dimensional time series, for which this smoothness is generally absent.

The dimension $d$ is estimated through a series of bootstrap tests for the eigenvalues $\lambda^{(K)}_i$ of the operator $K(\cdot,\cdot).$ Alternative approaches, which will not be pursued here, include identifying a significant drop in the eigenvalues, minimizing forecast errors~\cite{Hyndman.2007} and information criteria approaches~\cite{Bai.2002,Bai.2007,Hallin.2007}. Each test works with a null hypothesis $H_0: \lambda^{(K)}_{d_0+1} = 0$ for a different value of $d_0$ (see Appendix~\ref{app:estimation} for details). We start with a $d_0$ that is too large, for example corresponding to an estimated eigenvalue $\hat{\lambda}^{(K)}_{d_0+1} $ that is extremely small. Then we test and every time $H_0$ does not get rejected we lower $d_0$ by one, until $H_0$ gets rejected. Our multiple testing procedure makes it more likely to overestimate the dimension of $\mathcal{M}$ than to underestimate it. This is preferable, because an overestimated dimension of ${\mathcal{M}}$ generally leads to a smaller denoising error than an underestimated dimension. The intuition behind this is as follows. When you project onto an erroneous direction of ${\mathcal{M}}$ the harm is relatively small, because the to-be-projected curves $Y_t(\cdot)$ do not have a component in that direction (apart potentially from a contribution of the noise curves). On the other hand, when ${\mathcal{M}}$ is mistakenly missing a direction, then an actually existing component of $Y_t(\cdot)$ is lost during projection and this generically causes a larger denoising error.

\subsection{Estimation of $\mathcal{M}_\varepsilon$} \label{sec:est_M_eps}
Estimation of the noise space $\mathcal{M}_\varepsilon$ starts straightforward by estimating the noise covariance through $\Sigma_{\varepsilon}(u,v) = \Sigma_{Y}(u,v) - \Sigma_{X}(u,v).$ Details can be found in Appendix~\ref{app:estimation}.

A bootstrap test to select $\hat{d}_\varepsilon$ is not available, unlike for selecting $\hat{d}.$ Instead, we use the FPCA-interpretation of the eigenvalues $\lambda_j^{(\varepsilon)}$ of $\Sigma_{\varepsilon}(\cdot,\cdot)$ as the variance of the $j$-th principal component. If the explained variance of the $(\hat{d}_\varepsilon+1)$-th eigenvalue falls below a certain threshold $\tau_{d_\varepsilon} \in (0,1),$ we take $\hat{d}_\varepsilon$ as the estimate for $d_\varepsilon.$ To be more precise, the condition that determines the estimator $\hat{d}_\varepsilon$ is
\begin{equation}
\frac{\hat{\lambda}_{\hat{d}_\varepsilon+1}^{(\varepsilon)}}{\sum_{j=1} \hat{\lambda}_{j}^{(\varepsilon)}} < \tau_{d_\varepsilon} \leq \frac{\hat{\lambda}_{\hat{d}_\varepsilon}^{(\varepsilon)}}{\sum_{j=1} \hat{\lambda}_{j}^{(\varepsilon)}} ,
\end{equation}
where the sum in the denominators is taken over all (positive) eigenvalues. Note that the estimated eigenvalues are assumed to be sorted in descending order. The threshold $\tau_{d_\varepsilon}$ is set by the practitioner.

A second sublety is that in what follows $\widehat{\Sigma}_\varepsilon(u,v) = \widehat{\Sigma}_Y(u,v) - \widehat{M}_0(u,v)$ is replaced by
\begin{equation}
\widehat{\Sigma}_\varepsilon^{+}(u,v) := \sum_{j=1}^{\hat{d}_\varepsilon} \hat{\lambda}^{(\varepsilon)}_j \hat{\phi}_j^{(\varepsilon)} (u) \hat{\phi}_j^{(\varepsilon)} (v), \label{eq:Sigma_eps_plus}
\end{equation}
where $\hat{\phi}_j^{(\varepsilon)} (\cdot)$ is the eigenfunction of $\widehat{\Sigma}_\varepsilon(\cdot,\cdot)$ associated with eigenvalue $\hat{\lambda}^{(\varepsilon)}_j.$ The reason is that due to estimation noise $\widehat{\Sigma}_\varepsilon(\cdot,\cdot)$ is not semi-positive definite. It is a known phenomenon (see \citetext{Chen.2021} and references therein) that even for small values of the estimation noise, the negative eigenvalues can remain relatively large. By removing these negative eigenvalues through \eqref{eq:Sigma_eps_plus} by hand, we reduce the estimation error and improve the denoising performance.

\subsection{Estimation of $\mathcal{M}_\parallel$ and $\mathcal{M}_\perp$} \label{sec:est_M_par_perp}
In order to estimate an orthonormal basis of $\mathcal{M}_\parallel,$ we use the fact that we have access to the covariance of the noise curves $\varepsilon_t^\parallel(\cdot)$ parallel to $\mathcal{M}$ via
\begin{equation}
\widehat{\Omega}_\parallel^{(\bpsi)}  = \int_\mathcal{I} \int_\mathcal{I}  \widehat{\bpsi}(u) \left( \widehat{\bpsi}(v)  \right)^T \widehat{\Sigma}_\varepsilon^{+}(u,v) \dd u \dd v  .
\end{equation}
This $(\hat{d} \times \hat{d})$ matrix is an estimate of the covariance matrix of the principal components of the curves $\varepsilon_t^\parallel(\cdot)$ with respect to the basis functions ${\bpsi}(\cdot)$ of ${\mathcal{M}}.$ Diagonalization leads to $\hat{d}$ orthonormal eigenvectors $\widehat{\Omega}_\parallel^{(\bpsi)}  \hat{\bvarepsilon}_j^\parallel = \hat{\lambda}_j^\parallel \hat{\bvarepsilon}_j^\parallel ,$ with $\hat{\lambda}_1^\parallel \geq \hat{\lambda}_2^\parallel  \geq \ldots \geq \hat{\lambda}_{\hat{d}}^\parallel \geq 0,$ from which an estimate for an orthonormal basis of $\mathcal{M}_\parallel$ can be constructed as
\begin{equation}
\widehat{\mathcal{M}}_\parallel = \text{span} \big( \hat{\phi}_1^{\parallel}(\cdot), \ldots , \hat{\phi}_{\hat{d}_\parallel}^{\parallel}(\cdot) \big), \qquad \text{where} \quad \hat{\phi}_j^{\parallel}(u) = \hat{\bvarepsilon}_j^\parallel \cdot \widehat{\bpsi} (u),
\end{equation}
for $j=1,2,\ldots,\hat{d}_\parallel.$ Here, analogous to $d_\varepsilon,$ the dimension of $\mathcal{M}_\parallel$ is selected via the conditions
\begin{equation}
\frac{\hat{\lambda}_{\hat{d}_\parallel+1}^{\parallel}}{\text{Tr}\left[ \widehat{\Sigma}_\varepsilon^{+}(\cdot,\cdot) \right]} 
< \tau_{d_\parallel} \leq
\frac{\hat{\lambda}_{\hat{d}_\parallel}^{\parallel}}{\text{Tr}\left[ \widehat{\Sigma}_\varepsilon^{+}(\cdot,\cdot) \right]} ,
\end{equation}
where $\tau_{d_\parallel} \in (0,1)$ is a threshold set by the practitioner. Note that the denominator in these conditions representents the total variance of the noise curves. The $j$-th estimated eigenvalue represents the variance of the noise mode $\hat{\phi}_j^\parallel(\cdot).$ It is therefore prudent to compare these eigenvalues to the total variance of the noise. If, for example, (nearly) all noise is in the part perpendicular to $\mathcal{M},$ you do not wish to take many parallel modes into account for the regression analysis of MISE-optimal denoising. By mutually comparing the eigenvalues of only the parallel part of the noise, there is the risk of overestimating the number of relevant parallel modes and thereby introducing too much estimation uncertainty, leading to a poor performance in terms of denoising.

Estimation of the perpendicular noise space $\mathcal{M}_\perp$ occurs as follows. First, the time series of the perpendicular noise curves is found through
$\hat{\varepsilon}_t^\perp(u) = \big( (I - \widehat{P}_\mathcal{M}) \hat{\phi}_i^{(\varepsilon)} \big) (u),$ where $\widehat{P}_\mathcal{M}(u,v) := \widehat{\bpsi}(u) \cdot \widehat{\bpsi}(v)$ and $\hat{\phi}_i^{(\varepsilon)}(\cdot)$ was defined in \eqref{eq:Sigma_eps_plus}. Based on this estimated time series the covariance $\widehat{\Sigma}_{\varepsilon^\perp}(\cdot,\cdot)$ is computed in the usual way. Let this estimated covariance have orthonormal eigenfunctions $\hat{\phi}_i^{\perp}(\cdot)$ with corresponding eigenvalues $\hat{\lambda}_i^\perp,$ sorted in descending order. The dimension of the perpendicular subspace $\mathcal{M}_\perp$ is then selected via the conditions
\begin{equation}
\frac{\hat{\lambda}_{\hat{d}_\perp+1}^{\perp}}{\text{Tr}\left[ \widehat{\Sigma}_\varepsilon^{+}(\cdot,\cdot) \right]} 
< \tau_{d_\perp} \leq
\frac{\hat{\lambda}_{\hat{d}_\perp}^{\perp}}{\text{Tr}\left[ \widehat{\Sigma}_\varepsilon^{+}(\cdot,\cdot)\right]} ,
\end{equation}
where $\tau_{d_\perp} \in (0,1)$ is a threshold set by the practitioner, and the functions $\hat{\phi}_1^{\perp}(\cdot), \ldots, \hat{\phi}_{\hat{d}_\perp}^{\perp}(\cdot)$ form an estimate of the basis of $\mathcal{M}_\perp.$

After having estimated the spaces $\mathcal{M}, \mathcal{M}_\varepsilon, \mathcal{M}_\parallel$ and $\mathcal{M}_\perp,$ estimating the denoised signal curves through MISE-optimal denoising is a straightforward application of \eqref{eq:Omega_perp_Omega_parallel_perp_def} and \eqref{eq:denoising_main_result}. For details, see Appendix~\ref{app:estimation_denoising}.

\subsection{Consistency of denoising procedure}
In the context of curve time series as considered in this paper, the aim of denoising is to reconstruct the signal curves $X_t(\cdot)$ from the observed curves $Y_t(\cdot),$ given the data of a curve time series $\left\{ Y_t(u) \right\}_{t=1}^n$ of length $n.$ A denoising procedure is consistent if the reconstruction $\widehat{X}_t^\text{den}(\cdot)$ as estimated from the observed data converges to the true signal curve $X_t(\cdot),$ possibly up to an irreducible observational noise component, as the length of the available time series $n$ increases. To be more precise, we call a denoising procedure consistent if
\begin{equation} \label{eq:consistency_definition}
\lim_{n \to \infty} \mise{den} = \mise{den}^\text{min},
\end{equation}
where $\mise{den}^\text{min}$ is defined as the theoretically achievable minimum of the denoising procedure. This definition applies to both MISE-optimal denoising, with a minimum defined in~\eqref{eq:MISE_minimum}, and to orthogonal denoising, with a minimum given by $\mise{ortho}^\text{min} = \text{Tr} \left[ \Omega_\parallel \right].$

\section{Simulation} \label{sec:simulation}
This section illustrates MISE-optimal denoising with an application to simulated data. Section~\ref{sec:results} analyzes its performance and compares it with orthogonal denoising. Section~\ref{sec:forecasting} uses both MISE-optimal and orthogonal denoising as a pre-processing step in a forecasting problem. We start by discussing the setup of the simulations.

\subsection{Setup} \label{sec:simulation_setup}
The setup of our simulation is inspired by \citetext{Bathia.2010} and \citetext{Chen.2022}. The data consists of a curve time series $\{Y_t(u)\}_{t=1}^n$ of $n$ observations. Each curve is defined on the interval $\mathcal{I} = [0,1]$ and represented on a grid of $200$ equidistant points on that interval. They are constructed as the sum of a signal and a noise curve, $Y_t(u) = X_t(u) + \varepsilon_t(u),$ where the signal curves are defined as
\begin{equation}
    X_t(u) = g_X(\lambda) \tilde{X}_t(u), \qquad \tilde{X}_t(u) = \sum_{j=1}^d \xi_{tj} \phi_j(u). \label{eq:sim_signal}
\end{equation}
Throughout our simulations, we work with three dynamical spaces $\mathcal{M}$ with respective dimensions $d=2,4$ and $6.$ The principal components $\xi_{tj}$ are simulated according to the VAR(1)-process in \eqref{eq:VAR1_assumption}. For $d=2$ this is specified by\footnote{See Appendix~\ref{app:var} for the specifications of the VAR(1)-processes for $d=4$ and $d=6.$}
\begin{equation} \label{eq:sim_VAR1_model_d2}
A = \left(
\begin{matrix}
0.14275022 & -0.61629756 \\
-0.4615736 & -0.49825869
\end{matrix}
\right),
\qquad
\Omega = \left(
\begin{matrix}
0.60977113 & -0.01529231 \\
-0.01529231 &  0.00121252
\end{matrix}
\right).
\end{equation}
The procedure by which we obtained these specific matrices $A$ and $\Omega$ for the VAR(1)-process is as follows. We start from the requirement that the lag-0 autocovariance of the principal components is diagonal, as stipulated by the Karhunen-Lo\`eve theorem. We choose
\begin{equation}
\Sigma_0 = \Ex{ \bxi_t \bxi^T_t } = \text{diag}(\lambda_1, \ldots, \lambda_d), \qquad \text{with}\quad 
\lambda_i = 0.2 \frac{i-1}{d-1} + 0.7 \frac{d-i}{d-1},
\end{equation}
for $i=1,2,\ldots,d.$ In the case $d=2$ this means $\lambda_1 = 0.7$ and $\lambda_2 = 0.2.$ The values of $A$ are drawn i.i.d.~from the standard normal distribution, after which the whole matrix is rescaled to make the absolute value of the largest eigenvalue equal to 0.8. This ensures stationarity of the VAR(1)-process, which requires that all eigenvalues of $A$ are within the unit circle. Finally, we use the first Yule-Walker equation to determine the covariance of the noise term, $\Omega = \Sigma_0 - A \Sigma_0 A^T.$ Since $\Omega$ constructed in this way is not necessarily semi-positive definite, this is checked explicitly. If it fails to be so, the process of randomly generating entries of the matrix $A$ is repeated until a semi-positive definite covariance matrix is obtained.

An alternative choice for our simulation setup would be to use the above procedure to generate a new VAR(1)-process every time we run our denoising algorithm. This has as an advantage that the results become independent of the particulars of the specific VAR(1)-processes for $d=2,4,6$ that we are using. However, it will also introduce an additional source of variance in our results. Since in most practical situations one is dealing with a single (unknown) underlying VAR-process, this source of variance is rather unrealistic and therefore we do not randomize over different VAR models in our data generating process.

We use $\phi_j(u) = \cos (2 \pi j u) + \sin (2 \pi j u)$ with $j=1,2,\ldots,d$ as orthonormal functions that form a basis of the dynamical space $\mathcal{M}.$ To conclude our discussion of the signal curves, we define a normalization pre-factor
\begin{equation}
g_X(\lambda) = \sqrt{\frac{1-\lambda}{\overline{\text{Var}}\left[\tilde{X}_t(\cdot)\right]}}, \qquad\text{with}\quad \overline{\text{Var}}\left[\tilde{X}_t(\cdot)\right] = \sum_{j=1}^d \Var{\xi_{tj}} = \sum_{j=1}^d \lambda_j .
\end{equation}
Here the parameter $\lambda = 1 - \overline{\text{Var}}\left[X_t(\cdot)\right] \in [0,1]$ plays the role of noise level, as defined in section~\ref{sec:noise_level}.

The other part from which the observed curves $\{Y_t(u)\}_{t=1}^n$ are constructed are the noise curves,
\begin{equation}
    \varepsilon_t(u) = g_\varepsilon(\lambda) \tilde{\varepsilon}_t(u), \qquad \tilde{\varepsilon}_t(u) = \sum_{j=1}^{d_\varepsilon} \frac{Z_{tj}}{a^{j-1}} \phi^{(\varepsilon)}_j(u),
\end{equation}
where $Z_{tj} \thicksim \text{N}(0,1)$ are mutually independent. Throughout this simulation we choose $a=1.5$ and $d_\varepsilon=8.$ This means that the noise mode with the smallest contribution, $\phi^{(\varepsilon)}_{d_\varepsilon}(\cdot),$ explains about 2\% of the total (integrated) variance of the noise curves. As an (orthonormal) basis for the noise space $\mathcal{M}_\varepsilon$ we use
\begin{equation}
    \phi_j^{(\varepsilon)}(u) = \left[ \cos \theta_j + \sin \theta_j \right] \cos (2 \pi j u) + \left[ \cos \theta_j - \sin \theta_j\right] \sin (2 \pi j u),
\end{equation}
where $j=1,2,\ldots,d_\varepsilon.$ This choice of parametrization allows us to control the orientation of the noise space with respect to the dynamical space. The angles between the basis vectors of $\mathcal{M}$ and $\mathcal{M}_\varepsilon$ with respect to the $L^2$-norm~\eqref{eq:L2_norm} are
\begin{equation}
\angle \big(\phi_i(u), \phi_j^{(\varepsilon)}(u)  \big) = \arccos \langle \phi_i(u), \phi_j^{(\varepsilon)}(u) \rangle = \delta_{i,j} \theta_j + (1-\delta_{i,j})\pi/2
\end{equation}
where $\delta_{i,j}$ is the Kronecker delta. By default we will take
\begin{equation} \label{eq:theta_def_simulation}
\theta_j = \left\{
\begin{matrix}
\pi/4 & \qquad j=1,2,\ldots, \text{min}(d,d_\varepsilon), \\
& \\
0 & \qquad j = \text{min}(d,d_\varepsilon) + 1, \ldots, d_\varepsilon.
\end{matrix}
\right.
\end{equation}
This means that each basis vector of $\mathcal{M}$ is at an angle of $\pi/4$ with one basis vector of $\mathcal{M}_\varepsilon,$ in a pairwise manner. All other pairs of basis vectors are perpendicular.

The normalization pre-factor for the noise curves is
\begin{equation}
g_\varepsilon(\lambda) = \sqrt{\frac{\lambda}{\overline{\text{Var}}\left[\tilde{\varepsilon}_t(\cdot)\right]}}, \qquad\text{with}\quad \overline{\text{Var}}\left[\tilde{\varepsilon}_t(\cdot)\right] = \sum_{j=1}^{d_\varepsilon}  \frac{1}{a^{2(j-1)}} = \frac{1 - \left(\frac{1}{a^2}\right)^{d_\varepsilon}}{1 - \frac{1}{a^2}}.
\end{equation}
This parametrization implies $\overline{\text{Var}}\left[\varepsilon_t(\cdot)\right] = \lambda.$ Since $\overline{\text{Var}}\left[Y_t(\cdot)\right] = \overline{\text{Var}}\left[X_t(\cdot)\right] + \overline{\text{Var}}\left[\varepsilon_t(\cdot)\right] = 1$, which confirms that $\lambda$ indeed plays the role of noise level $\overline{\text{Var}}\left[\varepsilon_t(\cdot)\right] / \overline{\text{Var}}\left[Y_t(\cdot)\right]$ as defined in section~\ref{sec:noise_level}. By controlling $\lambda,$ we control the noise level in the simulated data. Another advantage of this parametrization is that $\overline{\text{Var}}\left[Y_t(\cdot)\right]$ is kept at a constant, enabling a more fair comparison of denoising performance for different values of the signal-to-noise ratio.

\subsection{Selection of tuning parameters}
The definition of the kernel $K(\cdot,\cdot)$ in \eqref{eq:K_definition} depends on a parameter $q \in \mathbb{N}^+,$ which is the largest lag of the autocovariances that determine $K(\cdot,\cdot),$ as well as the coefficients $c_1, \ldots, c_q.$ It has been reported in differents contexts \cite{Bathia.2010,Lam.2011,Chen.2022} that estimation of the dynamical space $\mathcal{M}$ is rather insensitive to the particular choice of $q>1,$ with the understanding that choosing $q$ too large leads to a larger finite-sample noise in the estimator $\widehat{K}(\cdot,\cdot).$ Throughout this paper, we take $q=2$ and $c_1=c_2=1,$ i.e.~$K(u,v) =  N_1 (u,v) + N_2 (u,v).$

The thresholds introduced in Sections \ref{sec:est_M_eps} and \ref{sec:est_M_par_perp} for estimating orthonormal bases of $\mathcal{M}_\varepsilon$, $\mathcal{M}_\parallel$ and $\mathcal{M}_\perp$ are set at $\tau_{d_\varepsilon} = \tau_{d_\parallel} = \tau_{d_\perp}=0.01.$ We have experimented with different values of the thresholds and observed that the results of our simulations are always essentially the same. When the noise level $\lambda$  is increased, the denoising performance seems to improve for larger values of the threshold. This makes sense, because the noise is relatively large and therefore it is easier to estimate $\Sigma_\varepsilon(\cdot,\cdot)$ more reliably. Furthermore, noise modes with relatively small eigenvalues do contribute notably when the noise level is relatively large. The above reasons make larger threshold values for large $\lambda$ both robust for estimation noise and noticable in the observed denoising performance. Finally, notice that when one of the thresholds is zero MISE-optimal denoising reduces to orthogonal denoising ($\widehat{X}_t^\text{opt}(u) = \widehat{X}_t^\text{ortho}(u)= Y_t^\parallel(u)$). We have verified this experimentally.

\subsection{Performance of the denoising procedure} \label{sec:results}
\begin{figure}[h]
 \makebox[\textwidth][c]{\includegraphics[width=1.0\textwidth]{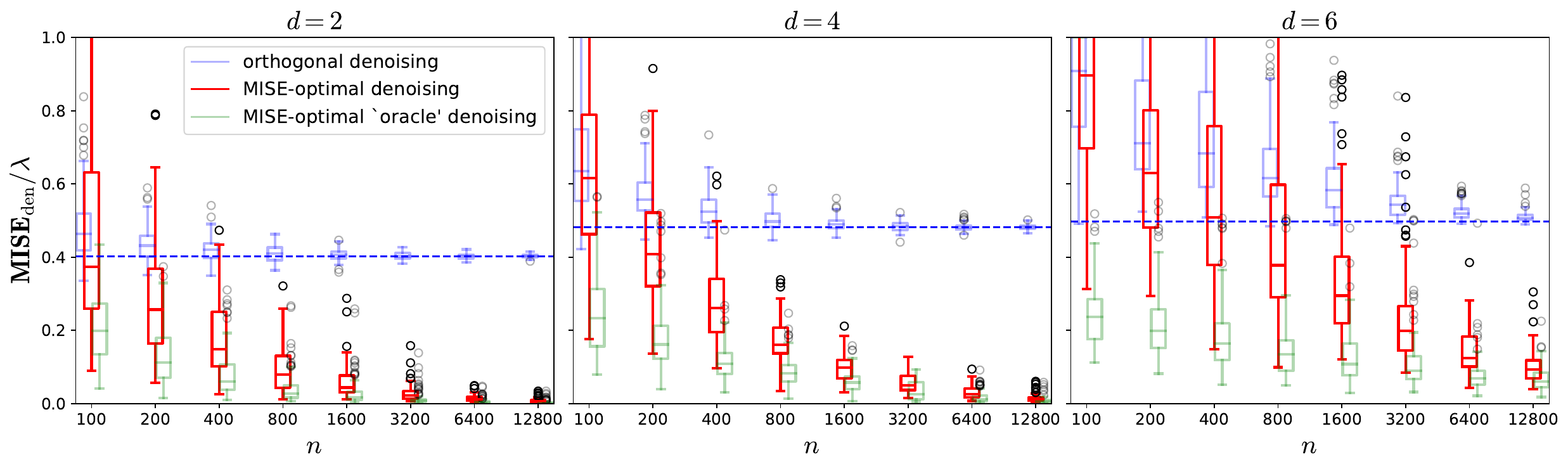}}
 \caption{Proportion of the (integrated) variance of the remaining noise after MISE-optimal denoising ({\it red}\,) as a function of the sample size $n.$ Same for orthogonal denoising ({\it blue}\,) and the ``oracle'' version ({\it green}\,). Also shown is the theoretical lower bound for orthogonal denoising ({\it blue, dashed line}).}
 \label{fig:asymptotics}
\end{figure}

We study the denoising performance of MISE-optimal and orthogonal denoising in Figure \ref{fig:asymptotics}, where the proportion of the (integrated) variance of the remaining noise after denoising, i.e.~$\mise{den}/\lambda,$ is plotted as a function of sample size $n.$ The noise level is fixed at $\lambda = 0.2$ and $\mise{den}$ is estimated in-sample.
For computational convenience the bootstrap estimate for $d$ is replaced by the true value. This is reasonable, as increasing the bootstrap sample size will bring the proportion of erroneous estimates to zero.
Every boxplot is based on 100 independent samples. Unless stated otherwise, we will use the same setup as described here.

Figure~\ref{fig:asymptotics} also shows the performance of an ``oracle'' version of the MISE-optimal denoising approach, for which both $\mathcal{M}$ and $\mathcal{M}_\varepsilon$ are known exactly and do not need to be estimated. This enables us to disentangle how the estimation of these subspaces on the one hand, and the regression step of MISE-optimal denoising on the other hand, contribute to the overall performance of the algorithm.

\begin{figure}[h]
 \makebox[\textwidth][c]{\includegraphics[width=1.0\textwidth]{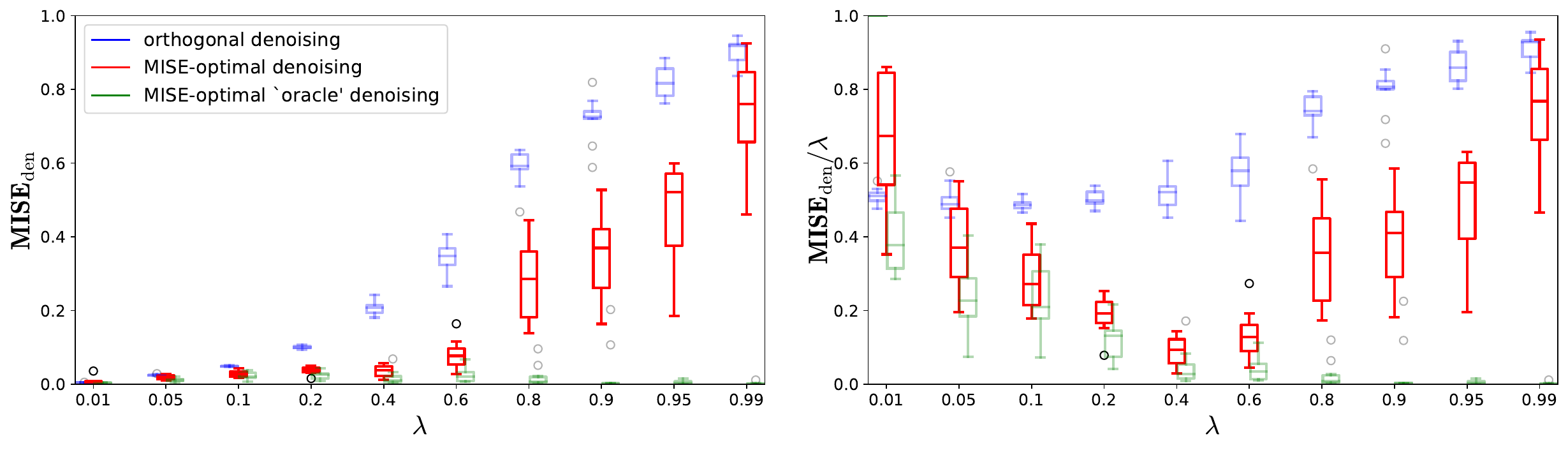}}
 \caption{MISE ({\it left panel}) and proportion of the (integrated) variance of the remaining noise ({\it right panel}) after MISE-optimal denoising ({\it red}\,) as a function of the noise level $\lambda.$ Same for orthogonal denoising ({\it blue}\,) and the ``oracle'' version ({\it green}\,).}
 \label{fig:denoising_function_of_lambda}
\end{figure}

The figure shows that MISE-optimal denoising is consistent, as all noise gets removed when $n$ tends to infinity. For $d=6$ convergence has not yet been achieved in the plot, but we verified it converges as well. Comparing MISE-optimal denoising with the ``oracle'' version, it is clear that for small $n$ most of the improvement comes from estimating $\mathcal{M}$ and $\mathcal{M}_\varepsilon,$ while for larger $n$ it comes from the regression step. For small $n$ the performance of the ``oracle'' version does not increase much as a function of $n$ (see $d=4$ and $d=6$), while for the same regime the performance of MISE-optimal denoising increases significantly as a function of $n.$ For larger $n$ both approaches converge at more or less the same speed.

It is worth mentioning that the performance of MISE-optimal denoising is particularly sensitive to a correct estimation of $d,$ the dimension of $\mathcal{M},$ for which (multiple) bootstrap tests were used. Almost all outliers in the boxplots for MISE-optimal denoising are due to erroneous estimates of $d.$

MISE-optimal denoising almost always (except for very small $n$ and large values of $d$) outperforms orthogonal denoising. Furthermore, orthogonal denoising does not lead to perfect denoising. To highlight this fact Figure \ref{fig:asymptotics} also exhibits a theoretical lower bound $\mise{ortho}^\text{min}$ (for details, see Appendix \ref{app:irreducible_noise}). This bound is due to the part of the noise that is parallel to $\mathcal{M}$ and therefore irreducible if you project orthogonally onto $\mathcal{M}.$

Figure \ref{fig:denoising_function_of_lambda} displays the denoising performances as a function of $\lambda,$ for fixed $n=800$ and $d=4.$ Only for very small noise levels ($\lambda=0.01$) our method is outperformed by the orthogonal denoising. This is explained by the fact that estimating $\mathcal{M}_\varepsilon$ is rather noisy for such small noise levels (note that the ``oracle'' version still outperforms orthogonal denoising). However, the same plot also shows that the impact of this underperformance is small. For such small noise levels the observed curves $Y_t(\cdot)$ are already very close to the signal curves and denoising does not change that.

\begin{figure}[h]
 \makebox[\textwidth][c]{\includegraphics[width=1.0\textwidth]{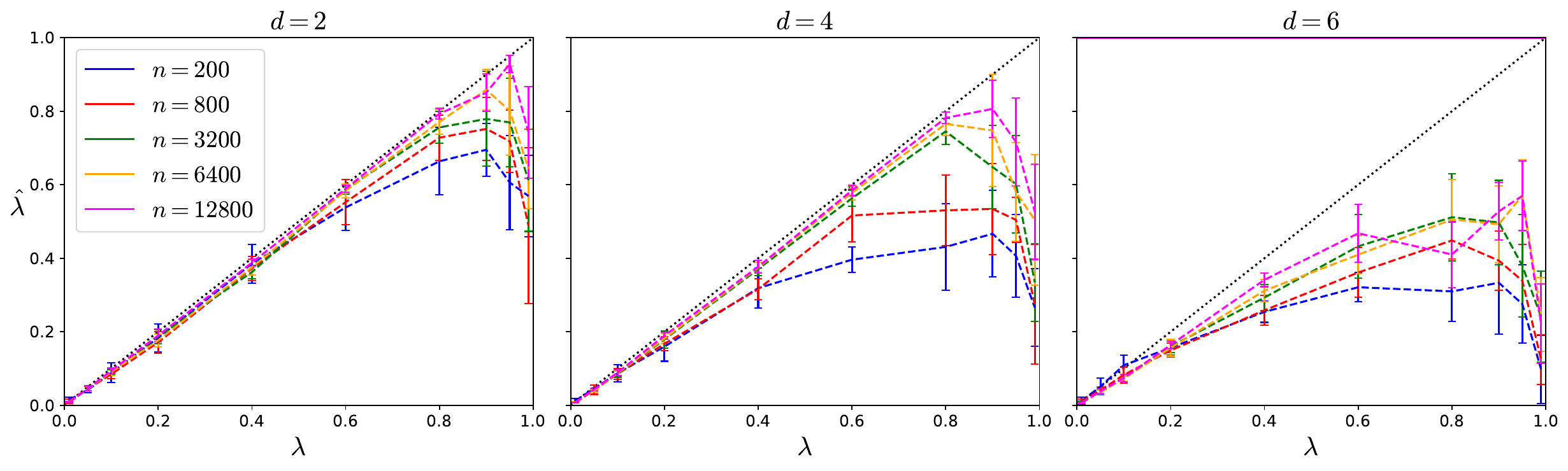}}
 \caption{Estimation of the noise level $\lambda$, based on \eqref{eq:reconstruct_lambda}. The error bars signify 1 sample standard deviation.}
 \label{fig:estimation_lambda}
\end{figure}

In Figure \ref{fig:estimation_lambda} we use \eqref{eq:reconstruct_lambda} to estimate the noise level $\lambda.$ Note that $\mise{opt}^\text{min}=0$ for this default simulation setup. For $d=2$ and $d=4$ we see that this approach leads to a consistent estimator of $\lambda$. The finite-sample bias leads to an underestimation of the noise level. An explanation might be the cutoffs we use to estimate the dimensions of $\mathcal{M}_\varepsilon$, $\mathcal{M}_\parallel$ and $\mathcal{M}_\perp.$ This is tantamount to neglecting a (small) portion of the noise. Similar to the results of Figure~\ref{fig:asymptotics}, convergence for $d=6$ is much slower.

Note that with $\lambda = \text{Tr} \left[\Sigma_\varepsilon \right]/ \overline{\text{Var}}\left[{Y}_t(\cdot)\right]$ we have an alternative approach to estimate $\lambda.$ Applying this method to our simulations, with $\widehat{\Sigma}_\varepsilon^{+}(u,v)$ as an estimator for $\Sigma_\varepsilon(u,v),$ leads to a second consistent estimator of $\lambda.$ However, the variance of this estimator is much larger (in particular for $d=4$ and $d=6$) and therefore we did not show it in the plots.

\begin{figure}[h]
 \makebox[\textwidth][c]{\includegraphics[width=1.0\textwidth]{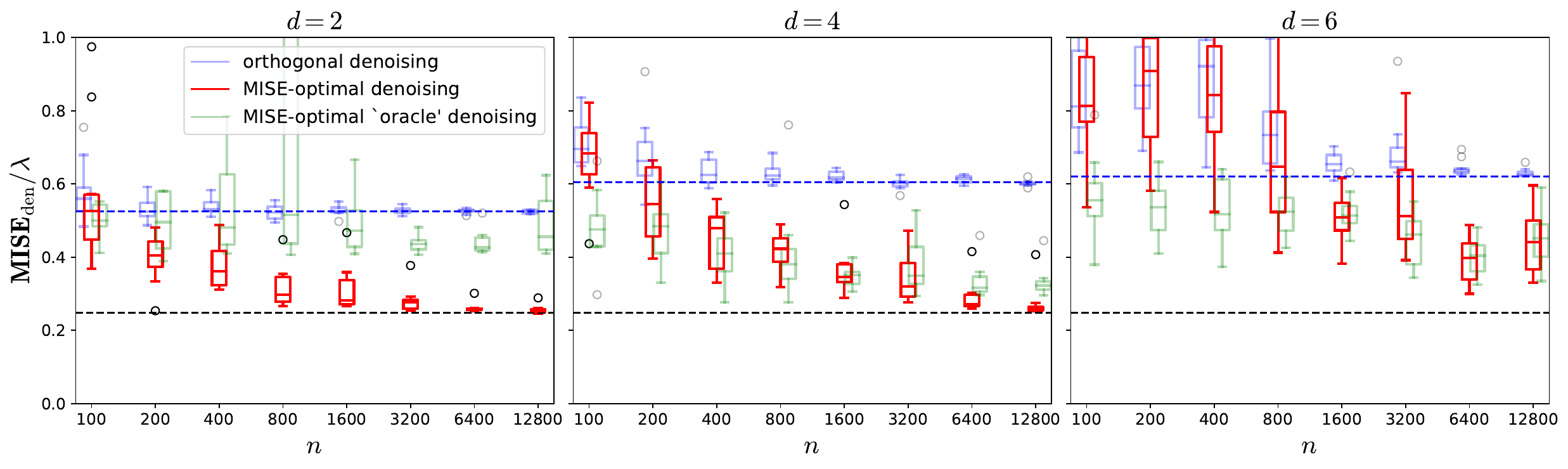}}
 \caption{Proportion of the (integrated) variance of the remaining noise after denoising. Similar to Figure \ref{fig:asymptotics}, except that now $\theta_2=0$ (cf.~\eqref{eq:theta_def_simulation}). Also shown is the theoretical lower bound for orthogonal denoising ({\it blue, dashed line}) and MISE-optimal denoising ({\it black, dashed line}).}
 \label{fig:asymptotics_zero_angle}
\end{figure}

In Figure~\ref{fig:asymptotics_zero_angle} we use exactly the same setup as in Figure~\ref{fig:asymptotics}, except that now $\theta_2 = 0$ (cf.~\eqref{eq:theta_def_simulation}). This means that $\mathcal{M} \cap \mathcal{M}_\varepsilon \neq \emptyset,$ i.e.~the signal and noise curves have a principal component direction in common. As a consequence, also MISE-optimal denoising has an irreducible noise component (see Appendix \ref{app:irreducible_noise} for details) given by $\mise{opt}^\text{min} = g_\varepsilon^2(\lambda)/a^2.$ This serves as a theoretical lower bound at the population level, and Figure~\ref{fig:asymptotics_zero_angle} clearly exhibits convergence towards this lower bound. This shows that also in the more general case of $\mathcal{M} \cap \mathcal{M}_\varepsilon \neq \emptyset$ MISE-optimal denoising is consistent.

Note that the ``oracle'' version with exact knowledge of $\mathcal{M}$ and $\mathcal{M}_\varepsilon$ is performing worse than MISE-optimal denoising, in particular for $d=2.$ This might seem counter-intuitive, because the ``oracle'' version has more information about the underlying data. It can be understood however by observing that $\theta_2=0$ reduces the dimension of $\mathcal{M}_\perp$ by one to $\text{dim} \, \mathcal{M}_\perp = d_\varepsilon - 1.$ Due to estimation noise the algorithm does not pick up this exact reduction and this makes $\widehat{\Omega}_\perp$ nearly singular, causing a poor denoising.

\begin{figure}[h]
 \makebox[\textwidth][c]{\includegraphics[width=1.0\textwidth]{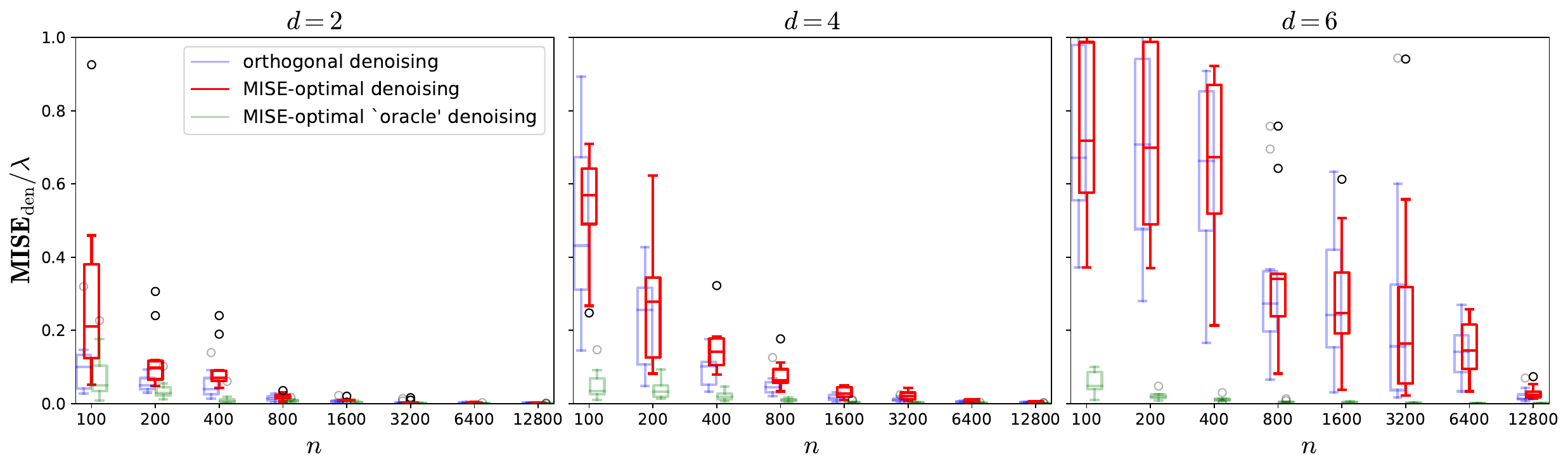}}
 \caption{Proportion of the (integrated) variance of the remaining noise after denoising. Similar to Figure~\ref{fig:asymptotics}, except that now $\theta_j = \pi/2$ for $j=1,\ldots,\text{min}(d,d_\varepsilon).$ This mean that $\mathcal{M}_\varepsilon$ is in the orthogonal complement of $\mathcal{M}.$}
 \label{fig:asymptotics_perp_angle}
\end{figure}

Finally, we investigate the case of all noise curves lying in the orthogonal complement of the signal curves: $\theta_j = \pi/2$ for $j=1,\ldots,\text{min}(d,d_\varepsilon).$ In the absence of parallel noise components, there is no dependent variable in the regression step and MISE-optimal denoising is equivalent to orthogonal denoising. Indeed, Figure~\ref{fig:asymptotics_perp_angle} shows that both methods have similar performances and convergence properties. MISE-optimal denoising even performs slightly worse than orthogonal projection, since due to estimation noise the result of the regression step is not a perfectly orthonogal projection onto $\mathcal{M}.$

\subsection{Forecasting} \label{sec:forecasting}
Forecasting of the simulated curve time series is possible by means of the FPCA framework and the extensive literature on forecasting of VAR models \cite{Lütkepohl.2005}. We focus on one-step-ahead forecasting the signal curve $X_t(u).$ In the hypothetical situation where the signal curves are accessible by direct observation, one can use the covariance operator $\Sigma_X(\cdot,\cdot)$ to find the Karhunen-Lo\`eve expansion \eqref{eq:finite_d_assumption}. The factor loadings $\bxi_{t}$ can be modelled with a VAR($p$) process,
\begin{equation} \label{eq:forecasting_VARp}
\bxi_t = \sum_{\ell=1}^p A_p \bxi_{t-\ell} + \be_t,
\end{equation}
where $\be_t \thicksim \text{IID} \left( {\bm 0}, \Omega\right),$ whose parameters can be least-square estimated. The one-step-ahead forecast of the signal curve at time $t$ is then given by
\begin{equation}
\widehat{X}_t^{(p)}(u) = \sum_{i=1}^d \hat{\xi}_{ti}^{(p)} \phi_i(u),
\end{equation}
where $\hat{\xi}_{ti}^{(p)} = \sum_{\ell=1}^p \widehat{A}_p \bxi_{t-\ell}$ is the optimal forecast of the factor loadings in the sense of minimizing the MSE of the forecast error \cite{Lütkepohl.2005}. To evaluate forecasting performance, we define a normalized mean integrated squared error
\begin{equation}
\Delta_{\text{F}} := \frac{
\overline{\text{E}} \big[ ( X_t(\cdot) - \widehat{X}_t^{(p)}(\cdot) )^2 \big]
}{
\overline{\text{E}} \big[ ( X_t(\cdot) )^2 \big]
}. \label{eq:MISE_forecasting}
\end{equation}
In reality the signal curves $X_t(\cdot)$ are usually not observable. The default workaround would be to use the observed curves $Y_t(\cdot)$ to compute the eigenfunctions of the Karhunen-Lo\`eve expansion. We have seen earlier that this introduces a bias due to the noise covariance. Since MISE-optimal denoising reduces the noise term, it is expected that using the denoised curves $\widehat{X}_t^\text{opt}(\cdot)$ instead will lead to a better forecasting performance.
\begin{table}[h]
\begin{center}
\begin{tabular}{ |c|c|c|c| } 
\hline
 $\lambda$ & 0.05 & 0.2 & 0.4 \\ \hline
 mean forecast & $0.998 \pm 1.16\mathrm{e}{-4}$ & $0.999 \pm 1.15\mathrm{e}{-4}$ & $0.999 \pm 9.98\mathrm{e}{-5}$ \\ 
 naive forecast & $2.335 \pm 5.83\mathrm{e}{-3}$ & $2.527 \pm 5.01\mathrm{e}{-3}$ & $2.935 \pm 6.00\mathrm{e}{-3}$ \\
 Karhunen-Lo\`eve & $0.623 \pm 1.56\mathrm{e}{-3}$ & $0.669 \pm 1.62\mathrm{e}{-3}$ & $0.767 \pm 2.52\mathrm{e}{-3}$ \\
MISE-optimal denoising & $0.614 \pm 1.59\mathrm{e}{-3}$ & $0.618 \pm 1.56\mathrm{e}{-3}$ & $0.628 \pm 1.61\mathrm{e}{-3}$ \\
 orthogonal denoising & $0.622 \pm 1.56\mathrm{e}{-3}$ & $0.654 \pm 1.53\mathrm{e}{-3}$ & $0.709 \pm 1.47\mathrm{e}{-3}$ \\ \hline
 theoretical lower bound & $0.617$ & $0.617$ & $0.617$ \\
 \hline 
\end{tabular}
\end{center}
\caption{Estimated forecast errors $\Delta_\text{F}$ based on a VAR(1) model of the factor loadings (including standard error of the mean). Estimations are based on a sample of size 200 with parameters $n=800$ and $d=4.$ The bottow row shows the theoretical lower bound of the forecast errors.}
\label{tab:forecasting_simulation}
\end{table}

In Table \ref{tab:forecasting_simulation} one-step ahead forecast errors $\Delta_\text{F}$ are computed in-sample, using five different approaches. On top of using the Karhunen-Lo\`eve expansion and the MISE-optimal denoised curves $\widehat{X}_t^\text{opt}(\cdot),$ we also use the orthogonally denoised curves $Y_t^\parallel(\cdot)$ and we introduce a ``mean forecast'' (using the mean of the observed curves as forecast) and a ``naive forecast'' (using the observed curve of the previous timestamp as forecast). Forecasting performance is also compared with the theoretical lower bound $g_X^2(\lambda) \text{Tr} \left[ \Omega \right]/(1-\lambda)$ of $\Delta_\text{F}$ due to the residual term in the VAR($p$) process defined in \eqref{eq:forecasting_VARp}. For the sake of simplicity, since we are mainly interested in differences in forecasting performance between different approaches, we take $p=1$ and use for MISE-optimal and orthogonal denoising knowledge of the true value of $d$ without estimating it. We find that MISE-optimal denoising outperforms the other forecasting strategies. The increase in performance compared to the Karhunen-Lo\`eve approach or the orthogonal-projection approach is not large in absolute terms, but it can be large in relative terms compared to the theoretical lower bound of the forecast error (in particular for $\lambda=0.4$).

\begin{figure}[h]
 \makebox[\textwidth][c]{\includegraphics[width=1.0\textwidth]{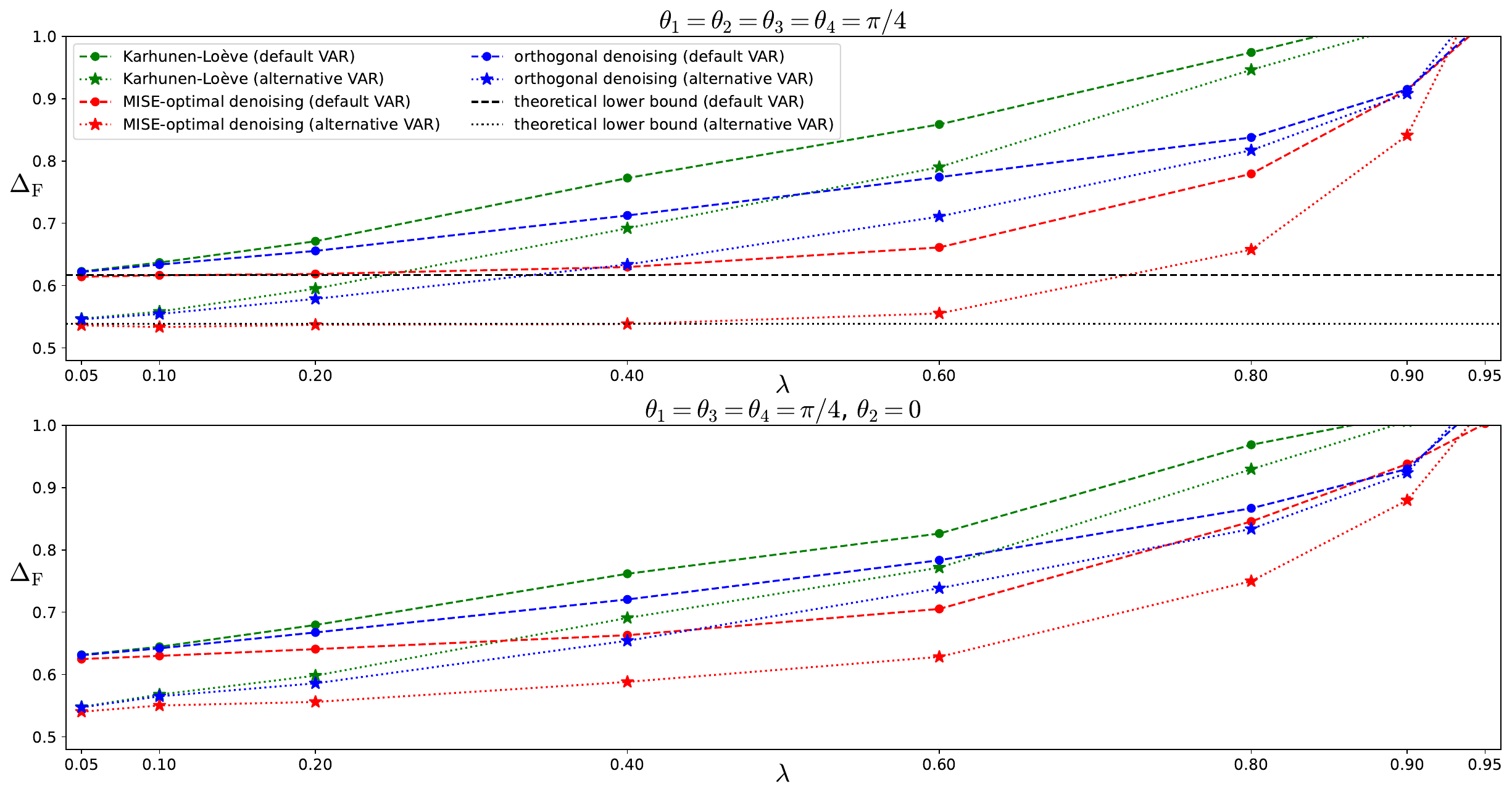}}
 \caption{Estimated forecast errors $\Delta_\text{F}$ based on a VAR(1) model of the factor loadings, as a function of the noise level $\lambda.$ Estimations are based on 50 simulations with parameters $n=800$ and $d=4.$ Standard errors are negligible. The lower panel is the case with irreducible noise.}
 \label{fig:forecasting}
\end{figure}

Figure \ref{fig:forecasting} gives more insights in the relevant factors affecting the forecasting performance $\Delta_\text{F}.$ In the top panel we compare forecasting by means for Karhunen-Lo\`eve, MISE-optimal denoising and orthogonal denoising, as a function of $\lambda.$ The angles between the dynamical and noise space are the default choice: $\theta_1=\theta_2=\theta_3=\theta_4=\pi/4.$ The figure shows results for the default VAR(1) model with $d=4,$ defined in \eqref{eq:sim_VAR1_model_d4}, and an ``alternative VAR(1) model'' with $d=4$ and eigenvalues of $A$ that are much closer to the unit circle. The bottom panel only differs in the fact that $\theta_2=0,$ meaning that there is an irreducible noise component.

There are several takeaways from Figure \ref{fig:forecasting}. First of all, forecasting based on MISE-optimal denoised signals outperforms the other approaches and the performance differences can be large (see for example $\lambda=0.60$). When the underlying VAR(1) model of the factor loadings has eigenvalues closer to the unit circle, all forecasting strategies have an improved performance. This makes sense, as larger eigenvalues mean a relatively stronger serial dependence of the time series (or, equivalently, relatively smaller residuals). Finally, comparing the two panels of Figure \ref{fig:forecasting}, we see that the relative performance increase due to MISE-optimal denoising is less in the presence of irreducible noise components. This is also as expected, since in this case the amount of noise that MISE-optimal denoising can reduce is more limited in a relative sense.

\section{Denoising empirical weather data} \label{sec:weather_data}
In this section MISE-optimal denoising is applied to empirical weather data. We consider hourly temperature measurements on 4015 consecutive days between 17 May 2012 and 14 May 2023. This means the curve time series has length 4015 and each curve consists of 24 measurements. The measurements were taken at a weather station in De Bilt, The Netherlands, and are made publicly available by the Royal Netherlands Meteorological Institute (KNMI).\footnote{\url{https://www.knmi.nl/nederland-nu/klimatologie/daggegevens}} The data coming from this particular weather station has been ``homogenized'' for eventual relocations of the weather station or changes in the measurement setup.

Before analyzing this data in the context of MISE-optimal denoising, three pre-processing steps are taken. Curve time series in the denoising methodology are assumed to be stationary. In order to make the empirical data more stationary, a correction is made for the seasonal trends in daily average temperature. At each day of the year an average daily temperature is estimated by means of a weighted Gaussian kernel with a bandwidth of 15 days and this average is subtracted from the data. Secondly, an hourly mean (the $\mu(u)$ in \eqref{eq:finite_d_assumption}) is estimated and subtracted from the data. Since this estimate turned out to be a smooth function, no additional smoothing was needed. Finally, the demeaned data is re-scaled such that it has a sample standard deviation of 1. This last step is strictly speaking not required by the proposed denoising method, but it makes interpreting the results more convenient. Note, for example, that the data now has the same variance as the observed curves of the simulated data in section \ref{sec:simulation}.

\begin{figure}[h]
 \makebox[\textwidth][c]{\includegraphics[width=1.0\textwidth]{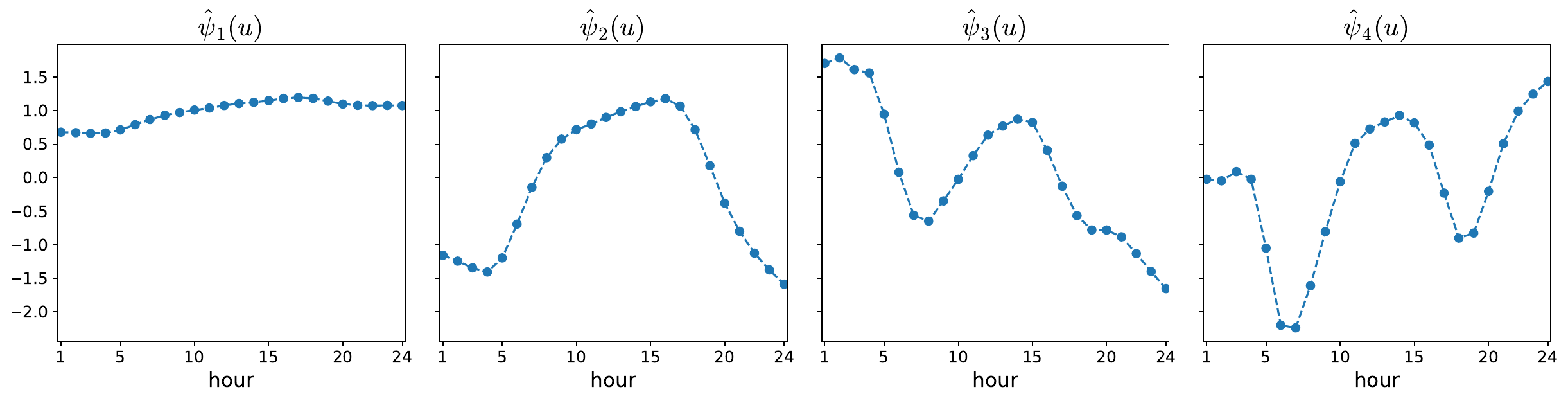}}
 \caption{Estimated basis of the dynamical space $\mathcal{M}.$}
 \label{fig:weather_M}
\end{figure}

\begin{figure}[h]
 \makebox[\textwidth][c]{\includegraphics[width=1.0\textwidth]{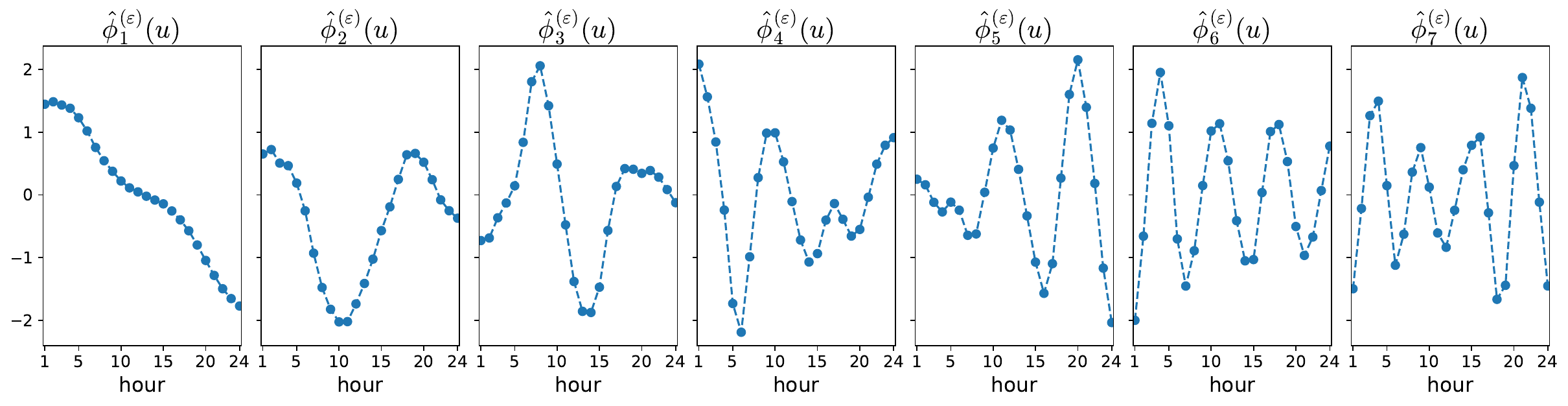}}
 \caption{Estimated basis of the noise space $\mathcal{M}_\varepsilon.$}
 \label{fig:weather_M_epsilon}
\end{figure}

Figures \ref{fig:weather_M} and \ref{fig:weather_M_epsilon} show the estimated orthonormal bases of the dynamical space $\mathcal{M}$ and the noise space $\mathcal{M_\varepsilon},$ respectively, with estimated dimensions $\hat{d} = 4$ and $\hat{d}_\varepsilon = 7.$ Hyperparameters for estimation are the same as for the simulated data in section \ref{sec:simulation} (here we use a bootstrap sample size of 100). The first basis function of the dynamical space, $\hat{\psi}_1(\cdot),$ is more or less constant and can be interpreted as the persisting daily average away from the seasonal average that was subtracted during pre-processing. Days of relative hot or cold weather are often clustered. The second basis function, $\hat{\psi}_2(\cdot),$ can be interpreted as the persisting day-night difference in temperature, away from the hourly mean $\mu(u)$ that was subtracted from the data. It suggests that ``diurnal air temperature variation'' (the difference in minimum/maximum temperature during one day) is persistent during a series of consecutive days, which is in agreement with meteorological research \cite{Cho.2020}. Interpreting the other two basis functions seems less trivial, apart from the fact that they seem to be odd and even, and that most variation occurs during sunrise and sunset. It should be noted though that the eigenvalue of $\hat{\psi}_1(\cdot)$ (recall that the basis functions are eigenfunctions of the operator $K(\cdot,\cdot)$) covers 96.20\% of the sum of the four eigenvalues, and the first two eigenvalues together cover 99.96\%. Finally, there does not seem to be a non-expert interpretation of the basis functions of the noise space. Perhaps not surprisingly, their frequency seems to be increasing gradually.

\begin{figure}[h]
 \makebox[\textwidth][c]{\includegraphics[width=1.0\textwidth]{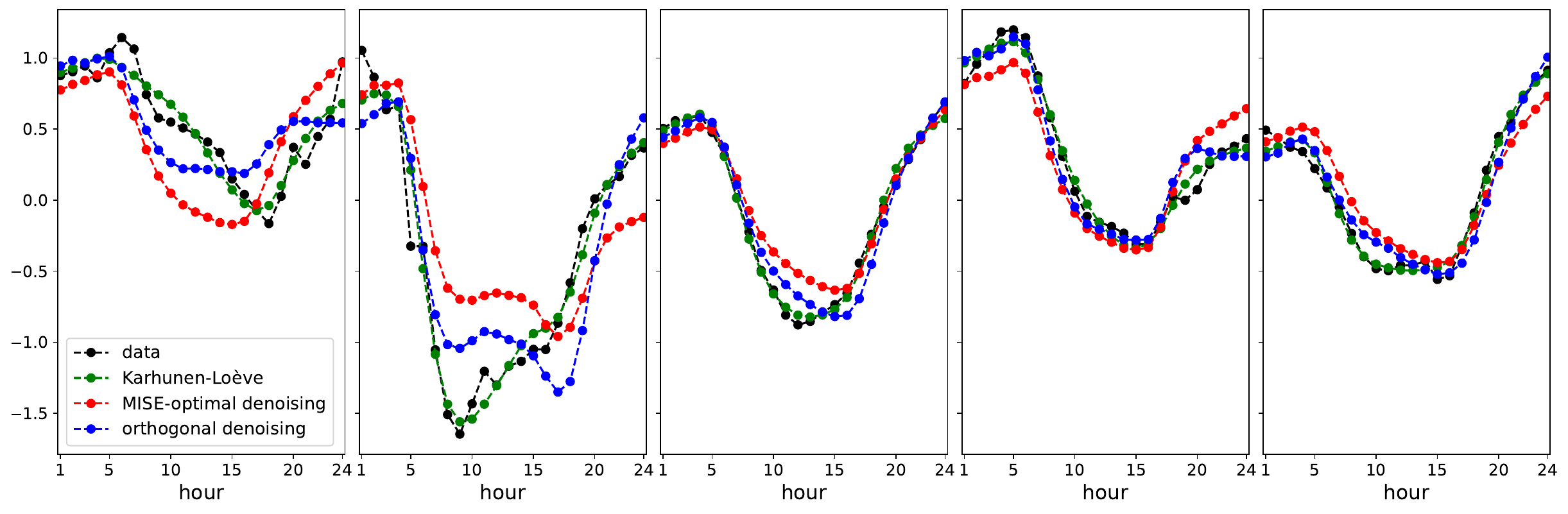}}
 \caption{MISE-optimal denoised curves for five consecutive days, alongside the original data, the orthogonally denoised curves and curves that are obtained by applying the Karhunen-Loève expansion to the observed data.}
 \label{fig:weather_denoising}
\end{figure}

We have applied MISE-optimal denoising to this data and find that $\hat{d}_\| = 3$ and $\hat{d}_\perp = 5.$ It is difficult to assess the validity of these estimates, as the data-generating process is unknown. However, it seems reassuring that the estimates are not at their theoretical minimum (0) and maximum ($\hat{d}_\varepsilon=7$) values. In Figure \ref{fig:weather_denoising} the MISE-optimal denoised curves are plotted for five consecutive days, alongside the original data, the orthogonally denoised curves and curves that are obtained by applying the Karhunen-Lo\`eve expansion directly to the observed data without denoising. For the latter we worked with 5 eigenfunctions, which together explain 98\% of the variance of the observed data. Note that this choice is rather arbitrary; for an increasing number of eigenfunctions the curves will converge to the original data. Also note that, unlike MISE-optimal and orthogonal denoising, this approach does not attempt to separate the dynamical, persistent part of the time series from the white-noise component.

Despite the absence of knowledge of the data-generating process, it is possible to estimate the proportion of the noise that has been reduced and the noise level in the original data (i.e.~$\lambda$). Consider the most general case, similar to the setup of Figure \ref{fig:asymptotics_zero_angle} of the simulations, where a part of the noise is irreducible. Furthermore, assume that MISE-optimal denoising has converged, i.e.~all reducible noise has been removed. Since here $\hat{d}=4$ and $n=4015,$ looking at Figure \ref{fig:asymptotics_zero_angle} this assumption seems reasonable but not entirely accurate. We have noted however that the first two basis functions of $\mathcal{M}$ cover 99.96\% of the total variance in the dynamical space, making $\mathcal{M}$ effectively two-dimensional and making the assumption more appropriate. As a consequence, we can estimate the MISE of the parts of the noise that remain after denoising and that have been removed by denoising:
\begin{subequations}
\begin{align}
\mise{opt}^{(\text{remaining})} & = \mise{opt}^\text{min} = \text{Tr}\left[ \Omega_\parallel \right] 
- \text{Tr}\left[ \left(\Omega_\perp\right)^{-1} \left(\Omega_{\parallel\perp}\right)^T \Omega_{\parallel\perp} \right] 
\,\widehat{=} \,\, 0.0759, \\
\mise{opt}^{(\text{removed})} & = 
\overline{\text{E}}\big[ ( Y_t(\cdot) - \widehat{X}_t^\text{opt}(\cdot) )^2 \big]
\,\widehat{=} \,\, 0.0884.
\end{align}
\end{subequations}
The proportion of the variance of the noise removed by MISE-optimal denoising is then
\begin{equation}
\frac{\mise{opt}^{(\text{removed})}}{\mise{opt}^{(\text{removed})} + \mise{opt}^{(\text{remaining})}}
\,\widehat{=} \,\, 0.538,
\end{equation}
and noise level is (cf.~\eqref{eq:reconstruct_lambda})
\begin{equation}
\lambda = \mise{opt}^{(\text{removed})} + \mise{opt}^{(\text{remaining})} \, \widehat{=} \,\, 0.164,
\end{equation}
where one should recall that the data has been re-scaled to have unit variance. In other words, 53.8\% of the variance of the noise is removed by MISE-optimal denoising. In the same fashion orthogonal denoising removes 27.2\% of the variance of the noise and gives the same estimate of $\lambda.$ We thus see that MISE-optimal denoising has a considerable advantage over orthogonal denoising in the case of this empirical dataset.

\begin{figure}[h]
 \makebox[\textwidth][c]{\includegraphics[width=1.0\textwidth]{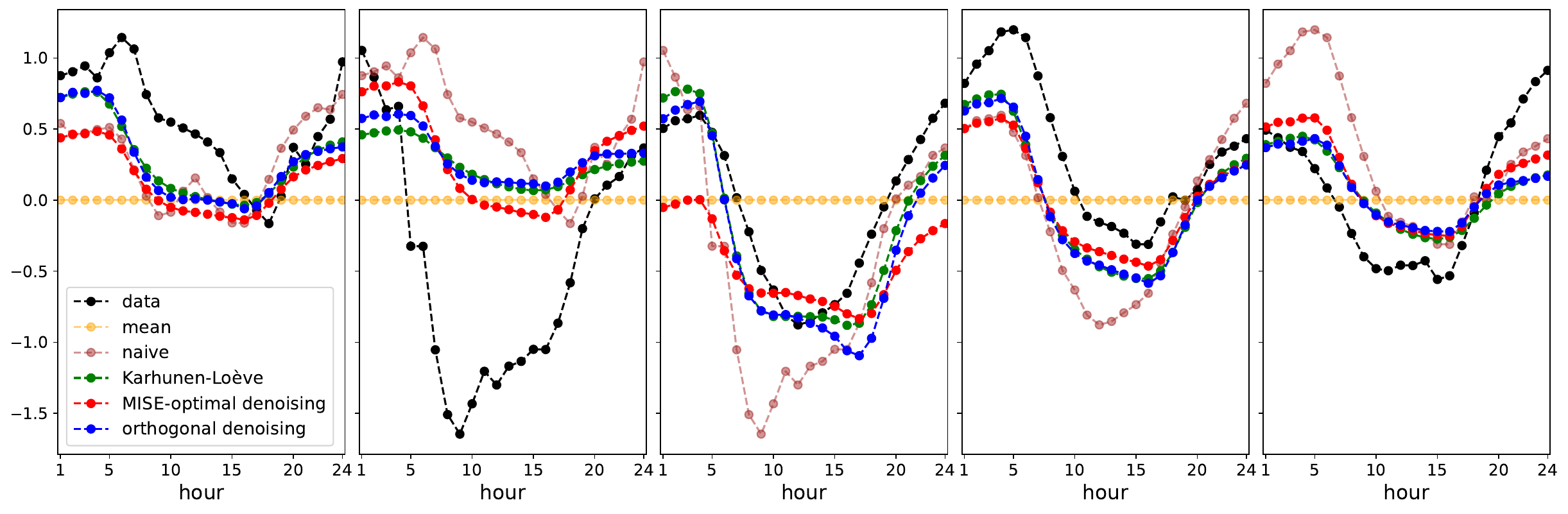}}
 \caption{One-day-ahead forecasts for five consecutive days based on five different methods, alongside the original data.}
 \label{fig:weather_forecasting}
\end{figure}

Finally, the five one-step-ahead forecasting approaches of section \ref{sec:forecasting} were {\it mutatis mutandis} applied to this dataset. Figure \ref{fig:weather_forecasting} shows the resulting forecasts for five consecutive days. In the absence of knowledge of the signal curves, assessing the forecasting performance is not straightforward. As a proxy, we replace the signal curves $X_t(\cdot)$ by the observed curves $Y_t(\cdot)$ in \eqref{eq:MISE_forecasting}. This leads to an extra contribution to the performance measure, coming from the noise curves, which is expected to be the same for all forecasting methods and therefore comparing forecasting performances is still possible.

\begin{table}[h]
\begin{center}
\begin{tabular}{ |c|c| } 
\hline
 mean forecast & $1.000$ \\ 
 naive forecast & $0.687$  \\
 Karhunen-Lo\`eve & $0.401$  \\
MISE-optimal denoising & $0.418$ \\
 orthogonal denoising & $0.406$ \\
 \hline 
\end{tabular}
\end{center}
\caption{Estimated forecast errors $\Delta_\text{F},$ with $X_t(\cdot)$ replaced by $Y_t(\cdot),$ for the empirical weather data, based on five forecasting methods and using a VAR(1) model for the factor loadings.}
\label{tab:forecasting_weather}
\end{table}

Table \ref{tab:forecasting_weather} shows that the forecasting performances of the Karhunen-Lo\`eve approach and the two denoising approaches are comparable, while clearly outperforming the mean and naive forecasts. The fact that MISE-optimal denoising clearly outperformed orthogonal denoising and the Karhunen-Lo\`eve approach, which essentially does not remove any noise, in terms of noise reduction, is not translated in a better forecasting performance. An explanation could be that the amount of reduced noise is only one of the many factors that influence forecasting performance, as Figure \ref{fig:forecasting} indicated for the simulated data. Considering that for the temperature data $\lambda \, \widehat{=} \,\, 0.164$ and the proportion of irreducible noise is higher than in the lower panel of Figure \ref{fig:forecasting}, it seems to make sense that the forecasting performances of the three approaches are close. Furthermore, the relative serial dependence within this curve time series could also play a role; if the serial dependence is small relative to $1/\lambda$, it will be difficult for denoising approaches to improve on the forecasting performance.

\newpage
\section{Conclusions and discussion}
By building on the DFPCA method of~\citetext{Bathia.2010}, this paper has introduced a method to disclose the structure of the noise space of functional time series and developed a MISE-optimal denoising procedure. By using simulated and real data, it was shown that this method outperforms DFPCA-based orthogonal denoising and can also be used as a pre-processing step to improve forecasting. We believe that this result can be seen as a ``blessing of dimensionality'' \cite{Gorban.2018}, since we can exploit the high dimensionality of the observed data to disentangle the dynamics and the noise in a curve time series.

An interesting open question stems from the restrictions imposed on the denoising operators we consider for minimizing the MISE. Only $Y_t(\cdot)$ is used to predict the underlying signal $X_t(\cdot).$ Adding information about the dynamics, for example $Y_{t-1}(\cdot),$ could potentially improve the denoising further.

We should also point out a hidden assumption about the noise. We apply the FPCA framework to the noise curves, assuming they can be approximated by a low-dimensional decomposition. It is unclear to what extend this assumption is valid for real data, or which impact a violation of this assumption has on MISE-optimal denoising.

Finally, let us mention a number of potential extensions and applications of MISE-optimal denoising. Although this paper considers denoising of functional time series, nowhere have we assumed anything about the smoothness of the curves of the time series. In other words, the ordering of the curves along the coordinate $u$ is irrelevant. Our method can therefore easily be extended to vector-valued, high-dimensional time series for which the factor model approach \cite{Lam.2012,Cubadda.2022,Dong.2022} is applicable.

To verify this explicitly, we have considered denoising of curve time series where the coordinate $u$ is shuffled. In the simulations in this paper the spatial coordinate is represented by a grid of $N=200$ equidistant points $u_1, u_2, \ldots, u_N$ on the interval $\mathcal{I}=[0,1].$ Given a curve time series $Y_t(\cdot),$ with a slight abuse of notation we can then introduce a ``shuffled time series'' $(\sigma Y)_t(u_i) = Y_t(u_{\sigma(i)}),$ where $\sigma$ is a permutation of the integers $\{1,2,\ldots,N\}.$ In the same fashion one can define $(\sigma X)_t(u_i)$ and $(\sigma \varepsilon)_t(u_i).$ We have verified numerically that MISE-optimal denoising applied to the shuffled time series leads to exactly the same denoising as when applied to the original curve time series. To be more concrete, we found that ${\widehat{(\sigma X)}}{\vphantom{\widehat{(\sigma X)}}^\text{opt}_t(u)} = {(\sigma \widehat X^\text{opt})}_t(u),$ or equivalently $\widehat X_t^\text{opt}(u) = \big( \sigma^{-1}{\widehat{(\sigma X)}}{\vphantom{\widehat{(\sigma X)}}^\text{opt}} \big)_t(u).$

Surface time series, which are relevant for e.g.~climate and environmental sciences \cite{Guillas.2010,Martinez.2020}, are another potential direction for extending the MISE-optimal denoising approach.

Another potential application is to use MISE-optimal denoising as a pre-processing step for functional linear regression. The hypothesis would be that using a denoised signal outperforms existing methods, which are essentially based on DFPCA and orthogonal denoising~\cite{Chen.2022,Chang.2023}.

\newpage
\bibliographystyle{oxford}
\bibliography{refs.bib}

\newpage
\appendix
\section{Curve time series, FPCA and DFPCA} \label{app:background}
This appendix contains some technical details about the definition of the curve time series we consider, FPCA and the DFPCA approach of \citetext{Bathia.2010}. It should be read as supplemental to sections \ref{sec:introduction} and \ref{sec:DFPCA}.

Let's start by making the definition of the curve time series (also called ``functional time series'') under consideration more precise. We have defined the noise term in \eqref{eq:curve_time_series} as the part of the observable curves $Y_t(\cdot)$ that does not have serial dependence. In particular, we will assume that $\varepsilon_t(\cdot)$ is a white noise sequence, as was done in, for example, \citetext{Bathia.2010,Chen.2022}. This means that $\Ex{\varepsilon_t(u)} = 0$ for all $t$ and any $u \in\mathcal{I},$ and that $\text{Cov} \left[ \varepsilon_t(u) , \varepsilon_s(v) \right]= 0$ for any $u, v \in\mathcal{I}$ and all $t \neq s$.  

Furthermore, we will assume that both the signal $X_t(\cdot)$ and the noise $\varepsilon_t(\cdot)$ are square integrable on the bounded interval $\mathcal{I},$ and that 
\begin{equation}
\int_\mathcal{I} \Ex{ X_t(u)^2 + \varepsilon_t(u)^2 }  \dd u 
\end{equation}
is finite and well-defined. Finally, we assume stationarity of the curve time series. This means that both
\begin{equation}
\mu(u) = \Ex{X_t(u)} \qquad\text{and}\qquad M_k(u,v) = \text{Cov} \left[ X_t(u), X_{t+k}(v) \right] ,
\end{equation}
where $k \in \mathbb{Z},$ are independent of $t.$

We apply the framework of FPCA to the signal curves $X_t(\cdot).$ As a starting point, we write the zero-lag autocovariance of the signal curves in terms of its spectral decomposition \eqref{eq:spectral_decomp_SigmaX}, where $\lambda_1 \geq \lambda_2 \geq \ldots \geq 0$ are the eigenvalues and $\left\{ \phi_i (\cdot) \right\}_{i=1}^\infty$ are the corresponding orthonormal eigenfunctions of the linear Hilbert-Schmidt operator
\begin{equation}
\Sigma_{X} : L^2(\mathcal{I}) \rightarrow L^2(\mathcal{I}), \qquad \Sigma_{X} (f) (u) = \int_\mathcal{I} \Sigma_{X}(u,v) f(v) \dd v. \label{eq:operator}
\end{equation}
The inner product on the Hilbert space is defined as
\begin{equation}
\left< f(\cdot), g(\cdot) \right> = \int_\mathcal{I} f(v) g(v) \dd v. \label{eq:L2_norm}
\end{equation}
Note that we are using the same symbol $\Sigma_{X}$ for the covariance function \eqref{eq:spectral_decomp_SigmaX} and the linear operator \eqref{eq:operator}, in order to keep notation simple. The existence of the expansion in \eqref{eq:spectral_decomp_SigmaX} is guaranteed by Mercer's theorem, which is applicable because $\Sigma_{X}(\cdot,\cdot)$ is a continuous, symmetric, non-negative definite kernel. 

The assumption of FPCA is that the (centered) curves $X_t(u) - \mu(u)$ can be approximately represented by a finite number of eigenfunctions,
\begin{equation}
X_t(u) \approx \mu(u) + \sum_{i=1}^m \xi_{ti} \phi_i(u) =: X_t^{(m)}(u) , \label{eq:FPCA_assumption}
\end{equation}
and that this approximation improves for increasing $m$, in the sense that
\begin{equation}
\lim_{m\to\infty} \Ex{\left( X_t(u) - X_t^{(m)}(u) \right)^2} = 0, \qquad \text{for all }u\in\mathcal{I}.
\end{equation}
Inspired by the framework of FPCA and similarly to \citetext{Bathia.2010}, we assume that the signal $X_t(\cdot)$ of the curve time series \eqref{eq:curve_time_series} under consideration is ``$d$-dimensional'', by which we mean that $\lambda_i = 0$ for $i >d.$ In other words, we assume that the exact identity \eqref{eq:finite_d_assumption} holds for a certain integer $d.$

In section \ref{sec:DFPCA} it was stated that almost any choice of the operator $K(\cdot,\cdot),$ defined in \eqref{eq:K_definition}, gives an eigenspace that coincides with the dynamical space $\mathcal{M}.$ In \citetext{Bathia.2010} it was namely proven that for any $k \geq 1$ for which the matrix $\Sigma_k = \Ex{\bxi_t \bxi_{t+k}^T}$ is full-rank, the operator
\begin{equation}
N_k (u,v) : = \int_\mathcal{I} M_k(u,z) M_k(v,z) \dd z \label{eq:Nk_definition}
\end{equation}
has exactly $d$ nonzero eigenvalues and their corresponding eigenfunctions span the dynamical space $\mathcal{M}.$ Since it can be argued \cite{Pena.1987,Pan.2008} that always for some $k \geq 1$, $\text{rank}(\Ex{\bxi_t \bxi_{t+k}^T}) = d$, this always provides an approach to estimating $d$ and $\mathcal{M}.$ As a consequence, the same properties hold for any operator of the form \eqref{eq:K_definition}, as long as there is a $k \in \{ 1, 2, \ldots q \}$ for which $c_k \neq 0 $ and $\Sigma_k$ is full-rank. For the sake of simplicity, we will assume throughout this paper that $\Sigma_k$ is of full rank for all $k \geq 1.$

The advantage of using $K(\cdot,\cdot)$ instead of $N_k(\cdot,\cdot)$ in the estimation procedure is that the former combines information about the dynamics of the underlying process from different lags. It is an interesting question what the optimal values for $q$ and $c_\ell$ in \eqref{eq:K_definition} are, in order to estimate $d$ and $\mathcal{M}.$ Intuitively, a small value for $q$ discards useful information stored in larger lags, while a large value for $q$ introduces more noise in the estimation coming from the larger lags. It has been reported however in different contexts~\cite{Bathia.2010,Lam.2011,Chen.2022} that the precise choice of $K(\cdot,\cdot)$ does not have much impact on the quality of the estimates of $d$ and $\mathcal{M}.$ We will therefore not investigate this question any further in the remainder of this paper.

\section{Vector autoregressive models \label{app:var}}
This appendix contains some technical details about vector autoregressive models relevant for our analysis, as well as a specification of the VAR(1) models used in the simulations of section \ref{sec:simulation}. See, for example, \citetext{Lütkepohl.2005} for more details about VAR models.

A general VAR(1) model is defined by
\begin{equation}
\vec{\xi}_t = \vec{\alpha} + A \vec{\xi}_{t-1} + \vec{e}_t,
\end{equation}
where $\vec{\xi}_t \in \mathbb{R}^d$, $\vec{\alpha} \in \mathbb{R}^d$  and the $d$-dimensional $\vec{e}_t \thicksim \text{IID} ( \vec{0}, \Omega ).$ The matrix $A$ is not necessarily symmetric, while the matrix $\Omega$ is. A VAR process is called stationary if 
\begin{equation}
\Ex{\vec{\xi}_t}, \qquad \Var{\vec{\xi}_t} \quad \text{and} \quad \bSigma_k := \text{Cov}\! \left[\vec{\xi}_t, \vec{\xi}_{t+k} \right]
\end{equation}
are independent of $t.$ If $A$ has all its eigenvalues within the unit circle, then it is a stationary process.

Let us now take the perspective of DFPCA as summarized in section \ref{sec:DFPCA}. Assume that we have chosen an operator $K(\cdot,\cdot)$ and that $\psi_1(\cdot), \psi_2(\cdot), \ldots ,\psi_d(\cdot)$ is the corresponding orthonormal set of eigenfunctions. This implies that
\begin{equation}
\mathcal{M}  =  \text{span}\{ \phi_1(u), \ldots ,\phi_d(u) \} =  \text{span}\{ \psi_1(u), \ldots ,\psi_d(u) \}.
\end{equation}
and, furthermore, that there is an orthogonal matrix $C$ such that $\bphi(u) = C \bpsi(u),$ where $\bphi(u) = \left(\phi_1(u),\ldots,\phi_d(u)\right)^T$ and $\bpsi(u) = \left(\psi_1(u),\ldots,\psi_d(u)\right)^T.$ Considering the two expansions of the demeaned signal curves in \eqref{eq:two_expansions}, this means that $\boldeta_t = C^T \bxi_t,$ where $\bxi_t$ was defined earlier as $\bxi_t = \left( \xi_{t1}, \ldots, \xi_{td} \right)^T$ and $\boldeta_t$ is defined similarly. We also have that ${\Sigma}_k^{(\eta)} = C^T {\Sigma}_k C,$ where the covariance matrices were defined below \eqref{eq:Mk_two_forms}. Recall that by the Karhunen-Lo\`eve theorem ${\Sigma}_0 = \text{diag} \left( \lambda_1, \ldots, \lambda_d \right)$ (see \eqref{eq:KarhunenLoeve_properties}).

In section \ref{sec:finding_noise_covariance} we devised a method for estimating the $\Sigma_\varepsilon(\cdot,\cdot),$ using the assumption \eqref{eq:VAR1_assumption} that the principal components $\bxi_t$ follow a VAR(1)-process. As a consequence of the assumption \eqref{eq:VAR1_assumption}, the random coefficients $\boldeta_t$ also follow a VAR(1) process,
\begin{equation}
\boldeta_t = A^{(\eta)} \, \boldeta_{t-1} + \be_t^{(\eta)}, \qquad \text{where} \quad \be_t^{(\eta)} \thicksim \text{IID} \left( \bm{0}, {\Omega^{(\eta)}} \right), \label{eq:VAR1_eta}
\end{equation}
and where $A^{(\eta)} = C^T A C$ and $\Omega^{(\eta)} = C^T \Omega C.$ The constant term is also absent here, because $\Ex{\boldeta_t} = \bm{0}$ as well. 

As a final step in section \ref{sec:finding_noise_covariance} we employ the Yule-Walker equations for the VAR(1) process of the $\boldeta_t$:
\begin{subequations}
\begin{align}
\Sigma_0^{(\eta)} & = A^{(\eta)} \Sigma_0^{(\eta)} \left( A^{(\eta)} \right)^T + \Omega^{(\eta)}, \label{eq:YW1} \\
\Sigma_k^{(\eta)} & = \Sigma_{k-1}^{(\eta)} \left( A^{(\eta)} \right)^T  \qquad \text{for } k>0. \label{eq:YW2}
\end{align}
\end{subequations}
The above equations are conventionally used to estimate the model parameters of the VAR(1) process. This so-called Yule-Walker estimator has the same asymptotic properties as the least-squares estimator, although for small samples it sometimes performs worse \cite{Lütkepohl.2005,Tjøstheim.1983}. We use the Yule-Walker equations differently, namely by ``reasoning backwards'' and computing $\Sigma_0^{(\eta)}$ through \eqref{eq:YW_reconstruction} from the lag-1 and lag-2 autocovariance matrices, which both can be estimated consistently.

\subsection*{VAR(1) models used in simulations}
As explained in section \ref{sec:simulation_setup}, throughout our simulations we work with three dynamical spaces $\mathcal{M}$ with dimensions $d=2,4$ and $6.$ The principal components $\xi_{tj}$ are simulated according to the VAR(1)-process in \eqref{eq:VAR1_assumption}. The model for $d=2$ is specified by in the main text. For $d=4$ we use
\begin{subequations}\label{eq:sim_VAR1_model_d4}
\begin{align}
A & = \left(
\begin{matrix}
-0.40475218 & 0.56881667 & -0.01251201 & -0.33319225 \\
0.36328118 & 0.23656237 &  0.17826015 & 0.47609812 \\
0.04062105 & -0.13439131 & -0.3596354 & -0.24931481 \\
-0.31412948 & 0.08911365 & -0.36549673 &  0.20076313
\end{matrix}
\right), \\
\Omega & = \left(
\begin{matrix}
0.39050087 & 0.06370578 & 0.0340153 & -0.10433378 \\
0.06370578 & 0.35412048 & 0.05387206 & 0.07341199 \\
0.0340153 & 0.05387206 & 0.2960237 & -0.02286662 \\
-0.10433378 & 0.07341199 & -0.02286662 & 0.06964716
\end{matrix}
\right).
\end{align}
\end{subequations}
For $d=6$ we use
\begin{subequations}
\small\begin{align}
A & = \left(
\begin{matrix}
0.37504966 & 0.08142893 & -0.07435684 & -0.03887785 & 0.25655029 & 0.25170869 \\
-0.14126954 & -0.19192149 & -0.0982056 & -0.37670302 & 0.16884435 & -0.38686508 \\
0.00451676 & -0.32514261 & -0.22975774 & 0.12353677 & 0.27258333 & 0.26566839 \\
0.44140703 & -0.08094657 & 0.05391765 & -0.09386828 & 0.03307928 & -0.14231888 \\
0.3419833 & -0.20556356 & 0.19934397 & 0.08967538 & 0.0027988 & 0.22842928 \\
0.01997925 & 0.10989784 & 0.29140585 & -0.007507 & 0.38542961 & 0.19185898
\end{matrix}
\right), \\
\Omega & = \left(
\begin{matrix}
 0.56177209 & 0.04343599 & -0.0262747 & -0.10676641 & -0.0826483 & -0.03922044 \\
0.04343599 & 0.46386005 & -0.02291322 & 0.01014503 & 0.05098028 & 0.02312988 \\
-0.0262747  & -0.02291322 & 0.36764981 & -0.00149778 &  -0.03508077 & 0.01351097 \\
-0.10676641 & 0.01014503 & -0.00149778 & 0.25032321  &  -0.11118431 & -0.00733749 \\
-0.0826483 & 0.05098028 & -0.03508077 & -0.11118431 & 0.15925535 & -0.02909279 \\
-0.03922044 & 0.02312988 & 0.01351097 & -0.00733749 & -0.02909279 & 0.09806406
\end{matrix}
\right)\!.
\end{align}
\end{subequations}
\normalsize

\section{M(I)SE-optimal denoising \label{app:denoising}}
This appendix provides a detailed derivation and discussion of the MISE-optimal denoising formula \eqref{eq:denoising_main_result}. It starts with a discussion of a much simpler case, namely MSE-optimal denoising for a two-dimensional time series. Then the $n$-dimensional generalization is considered, after which we discuss the case of functional time series. Because all cases are similar in essence, starting with more simple situations provides further insights in MISE-optimal denoising.

\subsection*{MSE-optimal denoising in $\mathbb{R}^2$}
Consider a stationary time series of a bivariate random variable $\vec{Y}_t = \vec{X}_t + \vec{\varepsilon}_t,$ with $\Ex{\vec{\varepsilon}_t}=0$ and $\Omega_\varepsilon = \Var{\vec{\varepsilon}_t}$ a (2x2)-matrix. As in the rest of this paper, assume that all persistence of the time series is contained in the ``signal'' part $\vec{X}_t$ and that $\vec{\varepsilon}_t$ is white noise. Furthermore, assume that the $\vec{X}_t$ lie in a one-dimensional ``dynamical space'' $\mathcal{M} = \text{span}(\vec{u}),$ where $\vec{u} \in \mathbb{R}^2$ is a unit vector. In other words, this is the case where $d=\dim \mathcal{M}=1.$ Note that this setup corresponds to the illustration of MISE-optimal denoising in Figure~\ref{fig:sketch_method}. We do not consider the case $d=2,$ because then necessarily $\mathcal{M}_\varepsilon \subset \mathcal{M}$ and (as we will see below) MSE-optimal denoising is not applicable.

Given an (observed) $\vec{Y}_t,$ the goal is to reconstruct the corresponding $\vec{X}_t$ in an MSE-optimal fashion. In other words, we would like to find a (linear) projection $P$ onto $\mathcal{M}$ that performs this reconstruction, $\vec{X}_t^\text{opt} = P\vec{Y}_t,$ and such that $\text{E}\big[||\vec{X}_t^\text{opt} - \vec{X}_t ||^2\big]$ is minimized. Observe that the noise can be uniquely decomposed into a part parallel to $\mathcal{M}$ and an orthogonal part, $\vec{\varepsilon}_t = \vec{\varepsilon}^{\,\,\parallel}_t + \vec{\varepsilon}^{\,\perp}_t.$ The idea behing MSE-optimal denoising is as follows. With knowledge of $\mathcal{M}$ and given a $\vec{Y},$ you can compute the orthogonal part of the noise and use this to reconstruct $\vec{X}.$ 

Because $P$ is a projection onto the dynamical space, $P\vec{X}_t=\vec{X}_t$ and thus $\vec{X}_t^\text{opt} = \vec{X}_t + P\vec{\varepsilon}_t.$ Let's write $\vec{\varepsilon}^{\,\perp}_t = \varepsilon^\perp_t \vec{v},$ where $\vec{v}\in\mathbb{R}^2$ is a unit vector perpendicular to $\vec{u}.$ Note that $\vec{v}$ is uniquely defined up to a sign. Since $P \vec{\varepsilon}^{\,\,\parallel}_t = \vec{\varepsilon}^{\,\,\parallel}_t$ and considering a specific form of the projection of the perpendicular noise, $P \vec{\varepsilon}^{\,\perp}_t = \alpha \varepsilon^\perp_t \vec{u},$ the objective of minimizing the MSE then translates into minimizing
\begin{equation}
\Ex{||\vec{X}_t^\text{opt} - \vec{X}_t ||^2} 
= \Ex{||P \vec{\varepsilon}_t ||^2} 
= \Var{\varepsilon^\parallel_t} + 2 \alpha \text{Cov}\!\left[ \varepsilon^\parallel_t, \varepsilon^\perp_t \right] + \alpha^2 \Var{\varepsilon^\perp_t} , \label{eq:cost_function_2d}
\end{equation}
where it was used that $\Ex{\vec{\varepsilon}_t}=0$ and $\varepsilon^\parallel_t$ was defined such that $\vec{\varepsilon}^{\,\,\parallel}_t = \varepsilon^\parallel_t \vec{u}.$ The (co)variances in \eqref{eq:cost_function_2d} are related to the original covariance matrix via an orthonormal basis transformation of the noise subspace $\mathcal{M}_\varepsilon,$
\begin{equation}
\left(
\begin{matrix}
\Var{\varepsilon^\parallel_t}  & \text{Cov}\!\left[ \varepsilon^\parallel_t, \varepsilon^\perp_t \right]  \\
\text{Cov}\!\left[ \varepsilon^\parallel_t, \varepsilon^\perp_t \right] & \Var{\varepsilon^\perp_t}
\end{matrix}
\right)
=
S^T \Omega_\varepsilon S, \qquad \text{where } S = \Big( \vec{u} \,\, \vec{v} \Big).
\end{equation}
The minimum of the convex cost function \eqref{eq:cost_function_2d} is given by
\begin{equation}
\hat{\alpha} = - \frac{\text{Cov}\!\left[ \varepsilon^\parallel_t, \varepsilon^\perp_t \right]}{\Var{\varepsilon^\perp_t}},
\end{equation}
leading to the MSE-optimal denoised signal
\begin{subequations}
\begin{align}
\widehat{X}_t^\text{opt}
& = \vec{X}_t + \hat{P} \vec{\varepsilon}_t \\
& = \vec{Y}^\parallel_t + \hat{P} \vec{\varepsilon}^{\,\perp}_t \\
& = P_\mathcal{M}\vec{Y}_t + \hat{\alpha} {\varepsilon}^\perp_t \vec{u} \\
& = \left[P_\mathcal{M} - \vec{u} \frac{\text{Cov}\!\left[ \varepsilon^\parallel_t, \varepsilon^\perp_t \right]}{\Var{\varepsilon^\perp_t}} \vec{v}^{\,T} \left( I - P_\mathcal{M} \right) \right] \vec{Y}_t, \label{eq:MOD_2d}
\end{align}
\end{subequations}
where $\vec{Y}^\parallel_t$ is the part of $\vec{Y}_t$ parallel to the signal subspace $\mathcal{M}$ and $P_\mathcal{M} = \vec{u} \vec{u}^{\,T}$ is the matrix of the orthogonal projection onto $\mathcal{M}.$ In the final line we used that $\varepsilon^\perp_t = \vec{v} \cdot \vec{\varepsilon}_t = \vec{v} \cdot \vec{\varepsilon}^{\,\perp}_t = \vec{v} \cdot \big(\vec{Y}_t - \vec{Y}^\parallel_t \big).$ The particular order of factors in \eqref{eq:MOD_2d}
was chosen with the prospect of generalizing this expression to higher dimensional time series. Finally, it should be noted that in essence the MSE-optimal denoising projection amounts to finding the MSE-optimal linear regression function of the conditional mean $\Ex{\varepsilon^\parallel_t | \varepsilon^\perp_t} = -\alpha \varepsilon^\perp_t,$ without an intercept because $\Ex{\vec{\varepsilon}_t}=0.$

The MSE-optimal denoising formula \eqref{eq:MOD_2d} exists provided that $\Var{\varepsilon^\perp_t} \neq 0.$ If this condition is not satisfied, the noise space $\mathcal{M}_\varepsilon$ is (effectively) one-dimensional and parallel to $\mathcal{M},$ i.e.~$\mathcal{M}_\varepsilon = \mathcal{M}.$ Intuitively it makes sense that a denoising projection is not possible in this situation. Given $\vec{Y}_t,$ since $\varepsilon^\perp_t=0$ we cannot use the part of the noise perpendicular to $\mathcal{M}$ to make a prediction for the noise parallel to $\mathcal{M}.$ The only remaining, sensible denoising option is subtracting a noise bias term from $\vec{Y}_t,$ but since we assumed $\Ex{\vec{\varepsilon}_t}=0$ this bias is zero.

Observe that when the noise parallel and perpendicular to the dynamical space $\mathcal{M}$ are uncorrelated the MSE-optimal projection reduces to the orthogonal projection onto $\mathcal{M}.$ This is as expected, since in this case $\varepsilon^\perp_t$ does not contain any information about $\varepsilon^\parallel_t$ and therefore cannot be used to project out (part of) $\varepsilon^\parallel_t.$

Also note that the minimized MSE is given by $\text{E}\big[||\hat{P} \vec{\varepsilon}_t ||^2\big] = \text{Var}\big[\varepsilon^\parallel_t\big] \left( 1 - \rho^2 \right),$ where $\rho$ is the correlation coefficient between $\varepsilon^\parallel_t$ and $\varepsilon^\perp_t.$ We get a perfect denoising (i.e.~zero error) in the special case of $\rho = \pm 1,$ which generally corresponds to the situation $\varepsilon^\parallel_t = \gamma \, \varepsilon^\perp_t$ for some finite $\gamma \neq 0.$ In other words, perfect denoising occurs when there is perfect correlation between the parallel and perpendicular parts of the noise. The noise space is then (effectively) one-dimensional, $\mathcal{M}_\varepsilon = \text{span}(\vec{w})$ with $\vec{w}$ a unit vector, and makes an angle with $\mathcal{M}$ that is given by $\tan^{-1}(1/\gamma).$ In this special case the MSE-optimal projection can take an alternative form $\hat{X}_t = \tilde{P} \vec{Y}_t,$ where
\begin{equation}
\tilde{P} = \Big( \vec{u} \,\, \vec{0}\, \Big) \left( A^T A \right)^{-1} A^T \qquad \text{and}\quad A = \Big( \vec{u} \,\, \vec{w} \Big).
\end{equation}
Assuming $\rho = \pm 1$ it is not difficult to show that the two projection formulas correspond, i.e.~$\hat{P} = \tilde{P}.$ From a geometric viewpoint, $\tilde{P}$ corresponds to a projection of $\vec{Y}_t$ parallel to $\mathcal{M}_\varepsilon$ onto $\mathcal{M}.$ This makes sense, as the noise is located on the line spanned by the vector $\vec{w}.$

Note that within this special situation there exist two particularly special cases. When $\gamma=0$ there is no noise parallel to $\mathcal{M},$ implying that $\vec{w} \perp \vec{u}.$ The projection corresponds to the orthogonal projection onto $\mathcal{M},$ meaning that $\tilde{P} = P_\mathcal{M},$ because $A^T A = I.$ The other special case is when $\gamma \to \infty.$ The noise then becomes parallel to $\mathcal{M},$ implying that $A^T A$ becomes a singular matrix and the projection formula for $\tilde{P}$ breaks down. This corresponds to the situation $\mathcal{M}_\varepsilon = \mathcal{M}$ which was discussed earlier and for which no MSE-optimal denoising exists.

\subsection*{MSE-optimal denoising in $\mathbb{R}^n$}
This section generalizes the situation of the previous section to a time series of an $n$-dimensional random variable $\vec{Y}_t = \vec{X}_t + \vec{\varepsilon}_t.$ Unless stated otherwise, the assumptions of the previous section are still valid. It is now assumed that the signal' $\vec{X}_t$ lies in a $d$-dimensional dynamical space $\mathcal{M}$ with $0 < d < n.$ The case where $d=n$ is not of interest, because in that case necessarily $\mathcal{M}_\varepsilon \subseteq \mathcal{M}$ and this renders MSE-optimal denoising impossible.

As in the previous section, the noise vector is (uniquely) decomposed as $\vec{\varepsilon}_t = \vec{\varepsilon}^{\,\,\parallel}_t + \vec{\varepsilon}^{\,\perp}_t,$ but now
\begin{equation}
\vec{\varepsilon}^{\,\,\parallel}_t = \sum_{i=1}^{d_\parallel} \tilde{\varepsilon}^{\,\parallel}_{t,i} \vec{u}_i, \qquad
\vec{\varepsilon}^{\,\perp} = \sum_{j=1}^{d_\perp} \tilde{\varepsilon}^{\perp}_{t,j} \vec{v}_j, 
\end{equation}
where the vectors $\vec{u}_1,\ldots,\vec{u}_{d_\parallel}$ and $\vec{v}_1,\ldots,\vec{v}_{d_\perp}$ form orthonormal basis of parallel and perpendicular noise spaces, $\mathcal{M}_\parallel$ and $\mathcal{M}_\perp,$ respectively. It is important to realize that the two bases combined do not necessarily form a basis of the total noise space $\mathcal{M}_\varepsilon.$ For example, in the case of perfect correlations between some parallel noise components $\tilde{\varepsilon}^{\,\parallel}_{t,i}$ and some perpendicular noise components $\tilde{\varepsilon}^{\perp}_{t,j},$ the total noise space will have a dimension $d_\varepsilon < d_\parallel + d_\perp.$ We will investigate this case further below. The current analysis is valid both in the absence of perfect correlations ($d_\varepsilon = d_\parallel + d_\perp$) and in the presence of perfect correlations ($d_\varepsilon < d_\parallel + d_\perp$).

The covariances of the parallel and perpendicular components of the noise can be computed from $\Omega_\varepsilon$ by means of coordinate transformations,
\begin{subequations}
\begin{align}
\Omega_\parallel & := \text{Cov}\! \left[ \tilde{\varepsilon}^{\,\parallel}_t,\tilde{\varepsilon}^{\,\parallel}_t \right] = U^T \Omega_\varepsilon U , \\
\Omega_\perp & := \text{Cov}\! \left[ \tilde{\varepsilon}^\perp_t,\tilde{\varepsilon}^\perp_t \right] = V^T \Omega_\varepsilon V , \\
\Omega_{\parallel\perp} & := \text{Cov}\! \left[ \tilde{\varepsilon}^{\,\parallel}_t,\tilde{\varepsilon}^\perp_t \right] = U^T \Omega_\varepsilon V ,
\end{align}
\end{subequations}
where
\begin{equation}
U = \Big( \vec{u}_1 \,\, \vec{u}_2 \, \ldots \, \vec{u}_{d_\parallel} \Big), \qquad V = \Big( \vec{v}_1 \,\, \vec{v}_2 \, \ldots \, \vec{v}_{d_\perp} \Big).
\end{equation}
Note that $\Omega_{\parallel\perp}$ is a matrix of dimensions $d_\parallel \times d_\perp.$

We again look for a projection $P$ that minimizes the MSE $\text{E}\big[||\vec{X}_t^\text{opt} - \vec{X}_t ||^2\big]$ for for this purpose we model the projection of the perpendicular noise component as
\begin{equation}
P \vec{\varepsilon}^{\,\perp}_t = \sum_{i=1}^{d_\parallel} \sum_{j=1}^{d_\perp} \alpha_{ij} \tilde{\varepsilon}^{\perp}_{t,j} \vec{u}_i = U \alpha \tilde{\varepsilon}^\perp_t. \label{eq:P_eps_perp_nd}
\end{equation}
The MSE then takes the form
\begin{subequations}
\begin{align}
\text{E}\big[||\vec{X}_t^\text{opt} - \vec{X}_t ||^2\big]
& = \Ex{||P \vec{\varepsilon}_t ||^2} \\
& = \sum_{i=1}^{d_\parallel} \Var{\tilde{\varepsilon}^{\,\parallel}_{t,i}} 
+ 2 \sum_{i=1}^{d_\parallel} \sum_{j=1}^{d_\perp} \alpha_{ij} \text{Cov}\!\left[ \tilde{\varepsilon}^{\,\parallel}_{t,i}, \tilde{\varepsilon}^{\perp}_{t,j} \right] \nonumber \\
& \qquad+ \sum_{i=1}^{d_\parallel} \sum_{j=1}^{d_\perp} \sum_{k=1}^{d_\perp} \alpha_{ij} \alpha_{ik} \text{Cov}\! \left[ \tilde{\varepsilon}^{\perp}_{t,j}, \tilde{\varepsilon}^{\perp}_{t,k}  \right]  \\
& = \text{Tr} \left[ \Var{\tilde{\varepsilon}^{\,\parallel}_t} \right] + 2 \, \text{Tr} \left[ \alpha^T \text{Cov}\!\left[ \tilde{\varepsilon}^{\,\parallel}_t, \tilde{\varepsilon}^\perp_t \right] \right] + \text{Tr} \left[ \alpha \, \text{Cov}\!\left[ \tilde{\varepsilon}^\perp_t, \tilde{\varepsilon}^\perp_t \right] \alpha^T \right] \\
& = \text{Tr} \left[ \Omega_\parallel \right] + 2 \, \text{Tr} \left[ \alpha^T \Omega_{\parallel\perp} \right] + \text{Tr} \left[ \alpha \, \Omega_\perp \alpha^T \right] , \label{eq:cost_function_dn}
\end{align}
\end{subequations}
This optimization problem is convex, since $\Omega_\perp$ is positive semi-definite and therefore $\text{Tr} \left[ \alpha \, \Omega_\perp \alpha^T \right] \geq 0$ for any nonzero matrix $\alpha.$ First order conditions are solved by $\hat{\alpha} = -\Omega_{\parallel\perp} \left( \Omega_\perp \right)^{-1},$ provided $\Omega_\perp$ is invertible. This results in an MSE-optimal denoised signal
\begin{subequations}
\begin{align}
\widehat{X}_t^\text{opt}
& = \vec{X}_t + \hat{P} \vec{\varepsilon}_t \\
& = \vec{Y}^\parallel_t + \hat{P} \vec{\varepsilon}^{\,\perp}_t \\
& = P_\mathcal{M}\vec{Y}_t + U \hat{\alpha} \tilde{\varepsilon}^\perp_t\\
& = \left[P_\mathcal{M} - U \, \Omega_{\parallel\perp} \left( \Omega_\perp \right)^{-1} V^T \left( I - P_\mathcal{M} \right) \right] \vec{Y}_t, \label{eq:MOD_nd}
\end{align}
\end{subequations}
where in the last line we used that $\tilde{\varepsilon}^{\perp}_{t,j} = \vec{v}_j \cdot \vec{\varepsilon}^\perp_t = \vec{v}_j \cdot \big( \vec{Y}_t - \vec{Y}^\parallel_t \big).$ This is the general MSE-optimal denoising result for $n$-dimensional time series.

Three special cases are worth to be mentioned here. First, there is the possibility that $\mathcal{M}_\varepsilon \subseteq \mathcal{M}.$ This means the absence of a noise component perpendicular to the dynamical space, $\vec{\varepsilon}_t = \vec{\varepsilon}^{\,\,\parallel}_t,$ and therefore the impossibility of MSE-optimal denoising. In the formulation of \eqref{eq:MOD_nd}, the matrix $\Omega_\perp$ is not defined in this situation. Secondly, there is the possibility that $\mathcal{M}_\varepsilon \subseteq \mathcal{M}^\perp.$ The noise is fully perpendicular to the dynamical space, $\vec{\varepsilon}_t = \vec{\varepsilon}^{\,\perp}_t,$ and MSE-optimal denoising reduces to orthogonal denoising: $\hat{P} = P_\mathcal{M}.$

At last, there is the already mentioned case of perfect correlations between parallel and perpendicular components of the noise. This implies $d_\varepsilon < d_\perp + d_\parallel.$ The general result \eqref{eq:MOD_nd} is still valid here, but there is an alternative formulation of MSE-optimal denoising. For those components of the parallel noise that are perfectly correlated with $\vec{\varepsilon}^{\,\perp}_t,$ it is now possible to project them out (completely) by projection in those directions parallel to the noise space $\mathcal{M}_\varepsilon.$

Let's illustrate this in the particular case where all parallel noise components are perfectly correlated with $\vec{\varepsilon}^{\,\perp}_t.$ In other words, consider the case where $\tilde{\varepsilon}^{\,\parallel}_t = \Gamma \tilde{\varepsilon}^\perp_t$ for some $(d_\parallel \times d_\perp)$-dimensional matrix $\Gamma.$ This is equivalent to saying that the noise space and the dynamical space do not have any directions in common, i.e.~$\mathcal{M}_\varepsilon \cap \mathcal{M} = \emptyset.$ Note that in this case $d_\varepsilon = \max (d_\parallel, d_\perp).$ If $\vec{w}_1, \vec{w}_2, \ldots, \vec{w}_{d_\varepsilon}$ is an orthonormal basis of $\mathcal{M}_\varepsilon,$ an alternative MSE-optimal projection equivalent to \eqref{eq:MOD_nd} is then given by
\begin{equation}
\tilde{P} = \Big( \vec{u}_1 \ldots \vec{u}_{d_\parallel} \, \vec{0} \ldots\vec{0} \Big) \left( A^T A \right)^{-1} A^T \qquad \text{where}\quad A = \Big( \vec{u}_1 \ldots \vec{u}_{d_\parallel} \, \vec{w}_1 \ldots \vec{w}_{d_\varepsilon} \Big).
\end{equation}
In this case the denoising is perfect in the sense that the minimized MSE is equal to zero.

\subsection*{MISE-optimal denoising in $L^2(\mathcal{I})$}
In this section details about the derivation of MISE-optimal denoising for curve time series is being discussed. It should be read as a supplement to section \ref{sec:denoising}, where the main steps of the derivation were presented. It can also be seen as a generalization of the previous sections of this appendix, from finite-dimensional time series to functional time series.

As soon as the parallel and perpendicular noise spaces are properly defined, as was done in section \ref{sec:denoising}, the derivation of MISE-optimal denoising is rather similar to the finite $n$-dimensional case of the previous section. The analogue of \eqref{eq:P_eps_perp_nd} was defined in \eqref{eq:denoising_linear_model}. Using that ${X}_t^\text{opt}(u) = X_t(u) + \varepsilon_t^\parallel(u) + (P \varepsilon_t^\perp)(u),$ the orthonormality of the bases of $\mathcal{M}_\parallel$ and $\mathcal{M}_\perp,$ as well as the fact that the noise curves have zero mean, one can easily find that
\begin{equation}
\mise{opt} = \text{Tr} \left[ \Omega_\parallel \right] + 2 \, \text{Tr} \left[ \alpha^T \Omega_{\parallel\perp} \right] + \text{Tr} \left[ \alpha \, \Omega_\perp \alpha^T \right],
\end{equation}
where $\Omega_\perp$ and $\Omega_{\parallel\perp}$ were defined in \eqref{eq:Omega_perp_Omega_parallel_perp_def} and
\begin{equation}
\Omega_\parallel := \int_\mathcal{I} \int_\mathcal{I}  \bphi^\parallel(u)\left( \bphi^\parallel(v)  \right)^T  \Sigma_{\varepsilon}(u,v) \dd u \dd v.
\end{equation}
As in the finite-dimensional case this is a convex optimization problem and the first-order conditions are solved by $\hat{\alpha} = -\Omega_{\parallel\perp} \left( \Omega_\perp \right)^{-1},$ provided $\Omega_\perp$ is invertible. The solution of the first-order conditions lead to the MISE-optimal denoising formula \eqref{eq:denoising_main_result}, the main result of this paper. Note that the curves $Y_t^\parallel(\cdot)$ in this expression can be conveniently expressed in terms of the basis vectors of $\mathcal{M}:$
\begin{equation}
Y_t^\parallel(u) :=  (P_\mathcal{M}Y_t )(u) = \int_\mathcal{I} \left[ \bpsi(u) \cdot \bpsi(v)\right] Y_t(v) \dd v \qquad \text{where}\quad P_\mathcal{M}(u,v) := \bpsi(u) \cdot \bpsi(v)
\end{equation}
is the operator for the orthogonal projection onto $\mathcal{M}.$

Similar to the finite-dimensional case, there are three special situations that need to be mentioned. In the absence of noise perpendicular to the dynamical space ($\mathcal{M}_\varepsilon \subseteq \mathcal{M}$) there is no MISE-optimal denoising possible. In the absence of noise parallel to the dynamical space, MISE-optimal denoising is equivalend to orthogonal denoising. And finally, in the case of perfect correlations ($d_\varepsilon < d_\parallel + d_\perp$) some parallel noise components can be removed completely. 

In the special case that all parallel noise components are perfectly correlated with the perpendicular noise (${\bvarepsilon}^{\parallel}_t = \Gamma {\bvarepsilon}^\perp_t$), one can project an observed curve $Y_t(\cdots)$ parallel to the noise space $\mathcal{M}_\varepsilon$ onto the dynamical space $\mathcal{M}$ and thereby remove all noise. The projection operator that achieves this is given by
\begin{equation}
\mathcal{P}_i (u,v) : = \phi_i (u) \sum_{j=1}^{d+d_\varepsilon} \left( \mathcal{A}^T \mathcal{A} \right)^{-1}_{ij} \tilde{\phi}_j(v)
\end{equation}
where 
\begin{equation}
\tilde{\bphi} (u) := \left( \phi_1 (u), \ldots, \phi_d (u), \phi^{(\varepsilon)}_1 (u), \ldots, \phi^{(\varepsilon)}_{d_\varepsilon}(u) \right)^T \quad \text{and} \quad \left( \mathcal{A}^T \mathcal{A} \right)_{ij} := \int_\mathcal{I} \tilde{\phi}_i(u) \tilde{\phi}_j(u) \dd u.
\end{equation}
The denoised signal is then
\begin{equation}
\tilde{X}_t^\text{opt}(u) = \sum_{i=1}^d \int_\mathcal{I} \mathcal{P}_i (u,v) Y_t(v) \dd v.
\end{equation}
At the population level this alternative denoising procedure is equivalent to our MISE-optimal denoising result \eqref{eq:denoising_main_result}. However, it turns out that at the level of finite-size samples this method performs less well, in the sense that it produces a larger MISE. The explanation is that $\mathcal{A}^T \mathcal{A} $ can quickly become nearly-singular due to estimation noise, causing large denoising errors.

\section{Estimation formulas for MISE-optimal denoising} \label{app:estimation}
This appendix contains the (rather standard) expressions for estimators needed for MISE-optimal denoising. It should be read as supplemental to section \ref{sec:estimation}.

\subsection*{Estimation of $\mathcal{M}$}
Defining an estimator for the operator $K(\cdot,\cdot)$ is straightforward,
\begin{equation}
\widehat{K}(u,v) := \sum_{\ell = 1}^p c_\ell \widehat{N}_\ell (u,v) , \label{eq:est_K}
\end{equation}
with
\begin{equation}
\widehat{N}_k(u,v) = \int_\mathcal{I} \widehat{M}_k(u,z) \widehat{M}_k(v,z) \dd z
\end{equation}
and where
\begin{equation}
\widehat{M}_k(u,v) = \frac{1}{n-k-1} \sum_{t=1}^{n-k} \left( Y_t(u) - \overline{Y}(u) \right) \left( Y_{t+k}(v) - \overline{Y}(v) \right),
\end{equation}
for $k=1,2,\ldots$ and where $\overline{Y}(u) = \tfrac{1}{n} \sum_{t=1}^n Y_t(u).$ Note that for $k=-1,-2,\ldots$ we have $\widehat{M}_k(u,v) = \widehat{M}_{-k}(v,u).$ The operator $\widehat{K}(\cdot,\cdot)$ has orthonormal eigenfunctions given by
\begin{equation}
(\widehat{K} \widehat{\psi}_j)(u) = \int_\mathcal{I} \widehat{K}(u,z) \widehat{\psi}_j(z) \dd z = \hat{\lambda}_j^{(K)} \widehat{\psi}_j(u) \label{eq:est_K_eigenspace}
\end{equation}
where $j=1,2,\ldots$ and $\hat{\lambda}_1^{(K)} \geq \hat{\lambda}_2^{(K)} \geq \ldots \geq 0.$ The estimated dynamical space is then given by $\widehat{\mathcal{M}} = \text{span} (\widehat{\psi}_1(u), \widehat{\psi}_2(u), \ldots, \widehat{\psi}_{\hat{d}}(u) ),$ where the dimension of the dynamical space is estimated via a bootstrap test (as was also done in \citetext{Bathia.2010,Chen.2022}). 

For this, we do multiple tests of the form $H_0: \lambda^{(K)}_{d_0+1} = 0$ with significance level $\alpha.$ The hypothesis is rejected when $\hat{\lambda}^{(K)}_{d_0+1} > c_\alpha,$ where $c_\alpha$ is the boundary of the rejection region. In order to apply the bootstrap procedure we define $\widehat{Y}_t(u) = \sum_{j=1}^{d_0} \widehat{\chi}_{tj} \widehat{\psi}_j(u),$ where
\begin{equation}
\widehat{\chi}_{tj} :=  \int_\mathcal{I} \left[ Y_t(u) - \overline{Y}(u)  \right] \widehat{\psi}_j(u) \dd u, \label{eq:est_principal_components}
\end{equation}
and then define $\hat{\varepsilon}_t(u) = Y_t(u) - \widehat{Y}_t(u).$ We then generate a bootstrap sample $Y_t^*(u) = \widehat{Y}_t(u) + \hat{\varepsilon}_t^*(u),$ where $ \hat{\varepsilon}_t^*(u)$ is drawn with replacement from the set $\left\{\hat{\varepsilon}_t(u)\right\}_{t=1}^n.$ Based on this bootstrap sample, we compute the $(d_0+1)$-th largest eigenvalue $\hat{\lambda}^{(K,*)}_{d_0+1}$ of the associated operator $\widehat{K}^*(u,v),$ similar to \eqref{eq:est_K} and \eqref{eq:est_K_eigenspace}. We repeat this for $B$ bootstrap samples and count how often $\hat{\lambda}^{(K)}_{d_0+1}  > \hat{\lambda}^{(K,*)}_{d_0+1}$ occurs. If this is more than $(1-\alpha)B$ times, we reject $H_0.$

\subsection*{Estimation of $\mathcal{M}_\varepsilon$} \label{app:est_M_eps}
Principal components $\widehat{\chi}_{tj}$ associated with the basis functions $\widehat{\bpsi}(\cdot)$ were defined in \eqref{eq:est_principal_components}. The lagged covariance matrices of the principal components $\eta_{tj}$ can now be estimated via
\begin{equation}
\widehat{\Sigma}_k^{(\eta)} = \frac{1}{n-k-1} \sum_{t=1}^{n-k} \widehat{\bchi}_{t} \widehat{\bchi}_{t+k}^T, \qquad\text{for k=1,2,\ldots.} 
\end{equation}
Note that for negative lags $k=-1,-2,\ldots$ we have $\widehat{\Sigma}_k^{(\eta)} = \left(\widehat{\Sigma}_{-k}^{(\eta)}\right)^T.$ By means of the Yule-Walker equations the estimator of the lag-$0$ covariance of the principal components $\eta_{tj}$ is then
\begin{equation}
\widehat{\Sigma}_0^{(\eta)} = \frac{1}{2} \left[ \, \widehat{\Sigma}_1^{(\eta)} \left( \widehat{\Sigma}_2^{(\eta)} \right)^{-1} \widehat{\Sigma}_1^{(\eta)} + \Big(\ldots\Big)^T \right],
\end{equation}
where on the $\dots$ there is a copy of the first term, such that $\widehat{\Sigma}_0^{(\eta)}$ is symmetric. Note that ${\Sigma}_0^{(\eta)}$ is necessarily symmetric, but due to finite-sample noise this is not the case for the estimator unless we explicitly symmetrize.

We are now in a position to estimate the covariance of the noise curves $\varepsilon_t(\cdot)$ through $\widehat{\Sigma}_\varepsilon(u,v) = \widehat{\Sigma}_Y(u,v) - \widehat{M}_0(u,v),$ where $\widehat{M}_0(u,v) = \widehat{\bpsi}(u)^T \widehat{\Sigma}_0^{(\eta)}  \widehat{\bpsi}(v)$ and 
\begin{equation}
\widehat{\Sigma}_Y(u,v) = \frac{1}{n-1} \sum_{t=1}^n \left( Y_t(u) - \overline{Y}(u) \right) \left( Y_t(v) - \overline{Y}(v) \right). \label{eq:Sigma_Y_hat}
\end{equation}
In parallel to how the dynamical space $\mathcal{M}$ was estimated, we can then estimate the noise space in terms of orthonormal eigenfunctions,
\begin{equation}
\widehat{\mathcal{M}}_\varepsilon = \text{span} \big( \hat{\phi}_1^{(\varepsilon)}(\cdot), \ldots , \hat{\phi}_{\hat{d}_\varepsilon}^{(\varepsilon)}(\cdot) \big), \qquad	\text{where} \quad \left( \widehat{\Sigma}_\varepsilon^+ \hat{\phi}_j^{(\varepsilon)} \right)\!(u) = \hat{\lambda}_j^{(\varepsilon)} \hat{\phi}_j^{(\varepsilon)} (u)
\end{equation}
and $\hat{\lambda}_1^{(\varepsilon)} \geq \hat{\lambda}_2^{(\varepsilon)} \geq \ldots \geq 0.$
First of all, note that in the above expression we have replaced $\widehat{\Sigma}_\varepsilon(\cdot,\cdot)$ by $\widehat{\Sigma}_\varepsilon^{+}(\cdot,\cdot),$ which was defined in \eqref{eq:Sigma_eps_plus}. See section \ref{sec:est_M_eps} for a discussion about this replacement.

\subsection*{Estimating the denoising operator} \label{app:estimation_denoising}
It is now straightforward to find the estimated MISE-optimal denoised signal curves,
\begin{equation}
\widehat{X}_t^\text{opt}(u) = \widehat{Y}_t^\parallel(u) - \int_\mathcal{I} \left( \hat{\bphi}^\parallel(u) \right)^T \widehat{\Omega}_{\parallel\perp} \left( \widehat{\Omega}_\perp \right)^{-1} \hat{\bphi}^\perp(v) \left( Y_t(v) - \widehat{Y}_t^\parallel(v) \right)  \dd v, \label{eq:est_denoising_main_result}
\end{equation}
where $\widehat{Y}_t^\parallel(u) =  (\widehat{P}_\mathcal{M}Y_t )(u) = \int_\mathcal{I} \left[ \widehat{\bpsi}(u) \cdot \widehat{\bpsi}(v)\right] Y_t(v) \dd v,$
\begin{equation}
\widehat{\Omega}_\perp  := \int_\mathcal{I} \int_\mathcal{I} \hat{\bphi}^\perp(u) \left( \hat{\bphi}^\perp(v)  \right)^T \widehat{\Sigma}_\varepsilon^{+}(u,v) \dd u \dd v  
\end{equation}
and
\begin{equation}
\widehat{\Omega}_{\parallel\perp} := \int_\mathcal{I} \int_\mathcal{I} \hat{\bphi}^\parallel(u) \left( \hat{\bphi}^\perp(v)  \right)^T \widehat{\Sigma}_\varepsilon^{+}(u,v) \dd u \dd v .
\end{equation}

\section{Dependence of the lag order for finding $\Sigma_0^{(\eta)}$} \label{app:VAR_order_dependence}
In Section \ref{sec:finding_noise_covariance} an estimation procedure for the autocovariance $\Sigma_\varepsilon(\cdot,\cdot)$ of the noise curves was derived by assuming that the principal components $\bxi_t$ of the signal curves $X_t(\cdot)$ follow a VAR(1) process. Here we investigate the impact of the assumption that the lag order is equal to 1. We first show that using a more general assumption of a VAR($p$) process with $p>1$ is possible and then analyze by numerical simulation the impact of choosing a larger lag order on MISE-optimal denoising.

If we assume that the principal components $\bxi_t$ follow a VAR($p$) process (instead of the VAR(1) process of \eqref{eq:VAR1_assumption}), the loadings $\boldeta_t$ also follow a VAR($p$) process:
\begin{equation}
\boldeta_t = \sum_{\ell = 1}^p A^{(\eta)}_\ell \, \boldeta_{t-\ell} + \be_t^{(\eta)}, \qquad \text{where} \quad \be_t^{(\eta)} \thicksim \text{IID} \left( \bm{0}, {\Omega^{(\eta)}} \right). \label{eq:VARp_eta}
\end{equation}
The Yule-Walker equations are then given by
\begin{subequations}
\begin{align}
    \Sigma_0^{(\eta)} & = \sum_{\ell=1}^p A^{(\eta)}_\ell \Sigma_\ell^{(\eta)} + \Omega^{(\eta)}, \\
    \Sigma_k^{(\eta)} & = \sum_{\ell=1}^p \Sigma_{k-\ell}^{(\eta)} \left( A^{(\eta)}_\ell \right)^T, \qquad \text{for}\quad k>0. \label{eq:YW_VARp_2}
\end{align}
\end{subequations}
We would like to express $\Sigma_0^{(\eta)}$ in terms of autocovariance matrices $\Sigma_k^{(\eta)}$ of non-zero lag, as the latter can be estimated using the proxy principal components $\chi_{tj}.$ For this, we need to eliminate the model parameter matrices $A^{(\eta)}_1, A^{(\eta)}_2, \ldots, A^{(\eta)}_p$ from the Yule-Walker equations. This is alike the standard Yule-Walker estimate, except that here we cannot include $\Sigma_0^{(\eta)}$ in our estimator.

We consider the Yule-Walker equations \eqref{eq:YW_VARp_2} for $k=p+1, p+2, \ldots, p+p,$ and write them as a system of linear equations for the model parameters:
\begin{equation}
    \left(
    \begin{matrix}
        \Sigma_{p+1}^{(\eta)} \\
        \Sigma_{p+2}^{(\eta)} \\
        \vdots \\
        \Sigma_{p+p}^{(\eta)} 
    \end{matrix}
    \right)
    =
    \left(
    \begin{matrix}
        \Sigma_{p}^{(\eta)} & \Sigma_{p-1}^{(\eta)} & \hdots & \Sigma_{1}^{(\eta)} \\
        \Sigma_{p+1}^{(\eta)} & \Sigma_{p}^{(\eta)} & \hdots & \Sigma_{2}^{(\eta)} \\
        \vdots & \vdots & & \vdots \\
        \Sigma_{p+p-1}^{(\eta)} & \Sigma_{p+p-2}^{(\eta)} & \hdots & \Sigma_{p}^{(\eta)}
    \end{matrix}
    \right)
    \left(
    \begin{matrix}
        \left( A^{(\eta)}_1 \right)^T \\
        \left( A^{(\eta)}_2 \right)^T \\
        \vdots \\
        \left( A^{(\eta)}_p \right)^T 
    \end{matrix}
    \right).
\end{equation}
This system can be solved for the model parameters $A^{(\eta)}_1, A^{(\eta)}_2, \ldots, A^{(\eta)}_p$ and subsequently we can use the Yule-Walker equation \eqref{eq:YW_VARp_2} with $k=p$ to find
\begin{equation}
    \Sigma_0^{(\eta)} = \left[ \Sigma_p^{(\eta)} - \sum_{\ell=1}^{p-1} \Sigma_{p-\ell}^{(\eta)} \left( A^{(\eta)}_\ell \right)^T \right] \left( \left( A^{(\eta)}_p \right)^T \right)^{-1}. \label{eq:Sigma0_eta_VARp}
\end{equation}
Hereby we have expressed $\Sigma_0^{(\eta)}$ in terms of quantities that can be estimated using the accessible principal components $\bchi_t$ defined in \eqref{eq:chi_definition}. This means that it is indeed possible to generalize the VAR(1) assumption to higher lag orders.

As an alternative to \eqref{eq:Sigma0_eta_VARp} we could have used the Yule-Walker equations \eqref{eq:YW_VARp_2} for $k=2,3,\ldots,p+1$ to arrive at the (almost) standard Yule-Walker estimates for the model parameters, which are then expressed in terms of the unknown $\Sigma_0^{(\eta)}.$ The Yule-Walker equation \eqref{eq:YW_VARp_2} for $k=1$ then leads to a highly non-linear equation for $\Sigma_0^{(\eta)},$ which could potentially be solved numerically. This has as an advantage that you need covariance matrices up to lag $p+1$ instead of lag $2p,$ but as disadvantage that there is no closed-form expression for $\Sigma_0^{(\eta)}.$ We do not pursue this option further.

Using the right-hand side of \eqref{eq:Sigma0_eta_VARp} to estimate $\Sigma_0^{(\eta)},$ one may question its estimation properties. A full analysis is beyond the scope of this paper, but it is worthwhile to point out that the regular Yule-Walker estimator for the model parameters of a VAR($p$) process has the same asymptotic properties as the least-squares estimator, although for small samples it sometimes performs worse \cite{Lütkepohl.2005,Tjøstheim.1983}. Furthermore, also for misspecified models the estimator is asymptotically optimal \cite{Dahlhaus.1996}. This gives reason to believe that \eqref{eq:Sigma0_eta_VARp} is a consistent estimator of $\Sigma_0^{(\eta)}.$

We investigate further the dependence of our denoising algorithm on the choice of the lag order $p$ by re-doing the simulations of Section \ref{sec:simulation}, but now assuming that the principal components $\bxi_t$ obey a VAR($p$) process with $p=1,2,3.$ The case $p=1$ is the default and analyzed in Section \ref{sec:simulation}. The cases $p=2$ and $p=3$ are new and have as main novelty that \eqref{eq:Sigma0_eta_VARp} is used to estimate $\Sigma_0^{(\eta)}.$ Apart from this, we use the same simulation setup as was used in Figure \ref{fig:asymptotics}.

\begin{figure}[h]
 \makebox[\textwidth][c]{\includegraphics[width=1.0\textwidth]{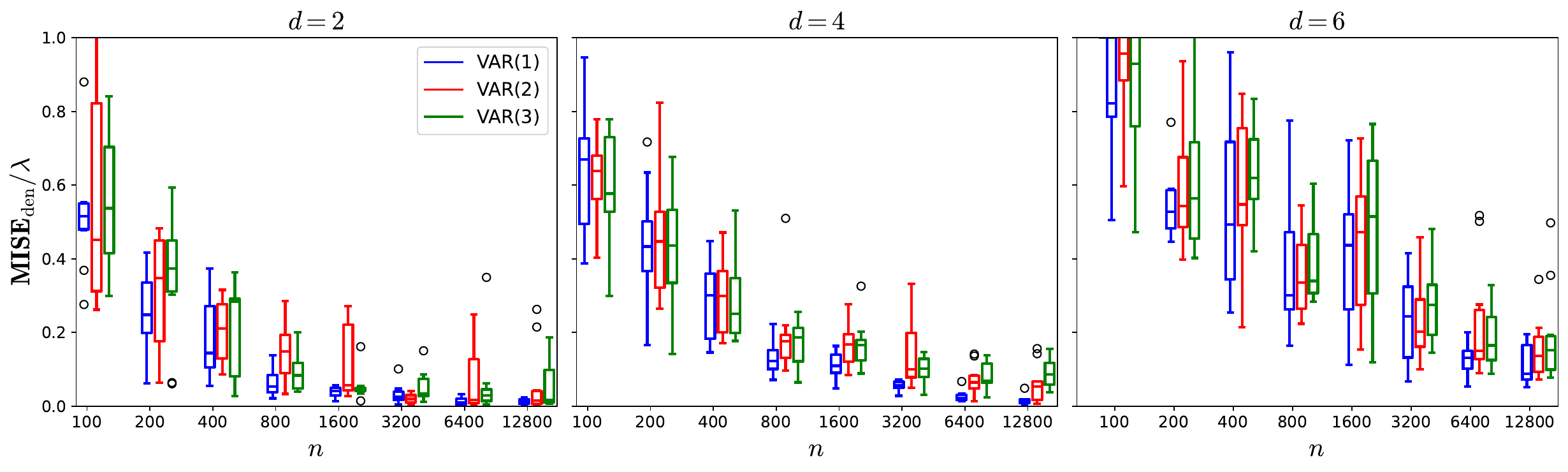}}
 \caption{Proportion of the (integrated) variance of the remaining noise after MISE-optimal denoising as a function of the sample size $n,$ for different lag orders ($p=1,2,3$) of the assumed underlying VAR($p$) process of the principal components $\bxi_t.$}
 \label{fig:VAR_order_dependence}
\end{figure}

The results are plotted in Figure \ref{fig:VAR_order_dependence}. Although the denoising performance for $p=1$ seems marginally better (in particular for small sample size $n$), the order of magnitude of the proportion of the (integrated) variance of the remaining noise and its asymptotic trend are similar for $p=1,2,3.$ This suggests that MISE-optimal denoising is insensitive to the choice of lag order $p$ of the assumed VAR($p$) model for the principal components.

Of course, making $p$ even larger will introduce more estimation noise, as the estimation of $\Sigma_0^{(\eta)}$ requires estimates of $\Sigma_k^{(\eta)}$ for $k=p+1,\ldots, p+p,$ and autocovariance matrices with larger lags will suffer more from finite-sample estimation noise. This seems to be the reason that the denoising performance for $p=1$ is marginally better. Furthermore, this is why we choose the VAR(1) assumption by default.

\section{Irreducible components of the noise curves} \label{app:irreducible_noise}
In this section we compute a theoretical lower bound for $\mise{opt},$ defined in \eqref{eq:MSE_objective}, in the context of the simulation of section~\ref{sec:simulation} in the case of $\theta_2=0$ (cf.~\eqref{eq:theta_def_simulation}). This means that $\mathcal{M} \cap \mathcal{M}_\varepsilon \neq \emptyset,$ i.e.~the signal and noise curves have a principal component direction in common. Since MISE-optimal denoising cannot filter out this component, the noise has in irreducible component and there will be a non-zero lower bound for the MISE. Let's focus on the simulation setup of section~\ref{sec:simulation}, and in particular on the basis functions of $\mathcal{M}$ and $\mathcal{M}_\varepsilon.$ They were defined as follows:
\begin{align}
\phi_j(u) & =  \cos (2 \pi j u) + \sin (2 \pi j u), \qquad j=1,\ldots,d, \\
\phi_k^{(\varepsilon)}(u) & = \left[ \cos \theta_k + \sin \theta_k \right] \cos (2 \pi k u) + \left[ \cos \theta_k - \sin \theta_k\right] \sin (2 \pi k u), \qquad j=1,\ldots,d_\varepsilon,
\end{align}
For simplicity we assume $d<d_\varepsilon,$ as is the case in all our simulations. If we decompose the noise modes as $\phi_k^{(\varepsilon)}(u) = \phi_k^{\parallel}(u) + \phi_k^{\perp}(u)$ with respect to $\mathcal{M},$ we find
\begin{subequations}
\begin{align}
\phi_k^{\parallel}(u) & = 
\left\{ 
\begin{matrix}
\cos (\theta_k) \phi_k(u), & \qquad k = 1,\ldots,d, \\
0 & \qquad k = d+1,\ldots,d_\varepsilon,
\end{matrix}
\right. \\
\nonumber \\
\phi_k^{\perp}(u) & = 
\left\{ 
\begin{matrix}
\sin (\theta_k) \left[ \cos (2 \pi k u) - \sin (2 \pi k u) \right], & \qquad k = 1,\ldots,d, \\
\phi_k^{(\varepsilon)}(u)  & \qquad k = d+1,\ldots,d_\varepsilon.
\end{matrix}
\right.
\end{align}
\end{subequations}
Note first that if $\theta_j \neq \pi/2$ for $j=1,\ldots,d,$ we have that $\mathcal{M}_\parallel = \mathcal{M}$ and we can use the functions $\phi_j(\cdot)$ as a basis for $\mathcal{M}_\|.$ Let's assume this is the case here. Furthermore, let's for now assume that $\theta_j \neq 0$ for $j=1,\ldots,d,$ which implies that $\text{dim}(\mathcal{M}_\perp) = d_\varepsilon$ and as an orthonormal basis for $\mathcal{M}_\perp$ we can use
\begin{equation}
\left\{
\begin{matrix}
\cos (2 \pi k u) - \sin (2 \pi k u) , & \qquad k = 1,\ldots,d, \\
\phi_k^{(\varepsilon)}(u)  & \qquad k = d+1,\ldots,d_\varepsilon.
\end{matrix}
\right.
\end{equation}
Eventually we are interested in the case $\theta_2=0,$ for which $\text{dim}(\mathcal{M}_\perp) = d_\varepsilon - 1$ and the second basis function in the above basis is removed. But for now we assume $\theta_2\neq 0.$

Using that
\begin{equation}
\Sigma_\varepsilon(u,v)  = \Ex{ \varepsilon_t(u) \varepsilon_t(v) } = g_\varepsilon^2(\lambda) \sum_{j=1}^{d_\varepsilon} \frac{1}{a^{2(j-1)}} \phi_j^{(\varepsilon)}(u) \phi_j^{(\varepsilon)}(v)
\end{equation}
we can find
\begin{equation}
\left( \Omega_\parallel \right)_{i,j}
 = \int_\mathcal{I} \int_\mathcal{I} \phi_i(u) \phi_j(v) \Sigma_\varepsilon(u,v) \dd u \dd v = \delta_{i,j} \frac{g_\varepsilon^2(\lambda)}{a^{2(j-1)}} \cos^2 (\theta_j)
\end{equation}
for $i,j=1,\ldots,d,$ and similarly
\begin{equation}
\left( \Omega_\perp \right)_{i,j} = \delta_{i,j} \frac{g_\varepsilon^2(\lambda)}{a^{2(j-1)}} \times \left\{
\begin{matrix}
\sin^2 (\theta_j) & \qquad j=1,\ldots,d, \\
1 & \qquad j = d+1,\ldots,d_\varepsilon,
\end{matrix}
\right.
\end{equation}
for $i=1,\ldots,d_\varepsilon,$ and
\begin{equation}
\left( \Omega_{\parallel\perp} \right)_{i,j} = \delta_{i,j} \frac{g_\varepsilon^2(\lambda)}{a^{2(j-1)}} \cos (\theta_j) \sin (\theta_j),
\end{equation}
where in the last expression $i=1,\ldots,d$ and $j = 1,\ldots,d_\varepsilon.$

The minimal value for the MISE is given in \eqref{eq:MISE_minimum}. Plugging in the above expressions we find that all terms in $\text{Tr}\left[ \Omega_\parallel \right]$ are canceled by the second trace and $\mise{opt}^\text{min} = 0.$ In this case perfect denoising is possible (at the population level).

Let's now focus on the case where $\theta_j=0$ for all $j\in \mathcal{J}_0 \subseteq \{1,\ldots,d\}.$ The term $\text{Tr}\left[ \Omega_\parallel \right]$ in $\mise{opt}^\text{min}$ is unaltered, but in the second trace the terms corresponding to the  $j\in \mathcal{J}_0$ are missing. This leads to 
\begin{equation} \label{eq:theoretical_lower_bound}
\mise{opt}^\text{min} = g_\varepsilon^2(\lambda) \sum_{j \in \mathcal{J}_0} \frac{1}{a^{2(j-1)}} .
\end{equation}
The case discussed in section~\ref{sec:results} is when $\mathcal{J}_0 = \{2\}.$

Denoising by means of orthogonal projection onto the dynamical space $\mathcal{M}$ also leads to an irreducible noise component, provided the noise space is not entirely inside the orthogonal complement of the dynamical space. Assuming the general simulation setup of section~\ref{sec:simulation} this lower bound can be computed analogously to how \eqref{eq:theoretical_lower_bound} was computed and is given by
\begin{equation} \label{eq:theoretical_lower_bound_ortho}
    \mise{ortho}^\text{min} = \text{Tr}\left[ \Omega_\parallel \right] = 
    g_\varepsilon^2(\lambda) \sum_{j \in \mathcal{J}_0} 
    \frac{\cos^2(\theta_j)}{a^{2(j-1)}} .
\end{equation}

\end{document}